%% file: Main.tex
\documentclass[format=acmsmall, review=false, screen=true]{acmart}

\usepackage{booktabs} 
\usepackage{bm}
\usepackage{amsmath}
\usepackage{multirow}
\usepackage{booktabs}
\usepackage{threeparttable}
\usepackage{ulem}

\usepackage{footnote}
\makesavenoteenv{table}

\usepackage{lscape}
\usepackage[ruled]{algorithm2e} 

\SetAlFnt{\small}
\SetAlCapFnt{\small}
\SetAlCapNameFnt{\small}
\SetAlCapHSkip{0pt}
\IncMargin{-\parindent}

\acmJournal{TIST}
\acmVolume{1}
\acmNumber{1}
\acmArticle{1}
\acmYear{2023}
\acmMonth{6}
\copyrightyear{2023}

\acmDOI{0000001.0000001}

\begin{document}
\title[Trustworthy Recommender Systems: An Overview]{Trustworthy Recommender Systems: An Overview}

\author{Shoujin Wang}
\affiliation{%
  \institution{University of Technology Sydney, RMIT University}
  \city{Sydney}
  \country{Australia}}
\email{shoujinwang@foxmail.com}

\author{Xiuzhen Zhang}
\affiliation{%
  \institution{RMIT University}
  \city{Melbourne}
  \country{Australia}
}
\email{xiuzhen.zhang@rmit.edu.au}

\author{Yan Wang}
\affiliation{%
  \institution{Macquarie University}
  \city{Sydney}
  \country{Australia}
}
\email{yan.wang@mq.edu.au}

\author{Huan Liu}
\affiliation{%
  \institution{Arizona State University}
  \city{Tempe}
  \country{United States}
}
\email{huan.liu@asu.edu}

\author{Francesco Ricci}
\affiliation{%
  \institution{Free University of Bozen-Bolzano}
  \city{Bolzano}
  \country{Italy}
}
\email{fricci@unibz.it}

\begin{abstract}

Recommender systems (RSs) aim to help users to effectively retrieve items of their interests from a large catalogue. 
For a quite long period of time, researchers and practitioners have been focusing on developing accurate RSs. 
Recent years have witnessed an increasing number of threats to RSs, coming from attacks, system and user generated noise, system bias. As a result, it has become clear that a strict focus on RS accuracy is limited and the research must consider other important factors, e.g., trustworthiness. 
For end users, a trustworthy RS (TRS) should not only be accurate, but also transparent, unbiased and fair as well as robust to noise or attacks. These observations actually led to a paradigm shift of the research on RSs: from accuracy-oriented RSs to TRSs. However, researchers lack a systematic overview and discussion of the literature in this novel and fast developing field of TRSs. To this end, in this paper, we provide an overview of TRSs, including a discussion of the motivation and basic concepts of TRSs, a presentation of the challenges in building TRSs, and a perspective on the future directions in this area. We also provide a novel conceptual framework to support the construction of TRSs.
\end{abstract}

%
%

\ccsdesc[500]{Information systems~Recommender systems}

%
%


\keywords{recommender systems, trustworthy recommendation}

\maketitle

\renewcommand{\shortauthors}{S. Wang et al.}

\input{paperbody.tex}

\end{document}

%% file: paperbody.tex
\section{Introduction}\label{Introduction}

We are living in the era of information explosion and digital economy, where the information overload problem has become increasingly important. As a matter of fact, today people usually make choices from massive and rapidly increasing catalogues of products and services (generally called \textit{items}), while consuming a large amount of time and resources to discriminate relevant from not-relevant items. To make informed choices and decisions in a more effective and efficient way, \textit{Recommender Systems (RSs)} have been introduced into almost every aspect of our daily life, work, business, study, entertainment and socialization~\cite{2022rsh,wang2021survey}. Today RSs are one of the most important and popular application areas of artificial intelligence (AI). Formally, RSs are software tools and techniques which provide suggestions on items which may be of interest to end users. According to McKinsey’s report\footnote{https://www.mckinsey.com/industries/retail/our-insights/how-retailers-can-keep-up-with-consumers}, 35\% of what customers purchase on Amazon and 75\% of what users watch on Netflix come from recommendations.

Since the release of the first RS, ``Tapestry'' in 1992, which was then termed collaborative filtering technique \cite{goldberg1992using}, RSs have flourished and achieved great success in the past 30 years. A series of RS approaches and models including content-based filtering, collaborative filtering, and hybrid approaches have been developed and deployed, most of which have achieved good recommendation performance. In recent years, benefiting from the advancement of machine learning, especially deep learning, more powerful and accurate RS models have been proposed~\cite{2022rsh,zhang2019deep}.

Although tremendous successes have been achieved in the RS area, the majority of existing research works has focused on the improvement of the system accuracy, while only some minor part of the literature has also taken some other important properties and values of RSs into account, such as diversity and novelty~\cite{CastellsHV15}. In fact, existing accuracy-oriented RSs have not properly considered end users in the broader context of human-machine interaction where other important factors are commonly considered as critical for the overall user experience~\cite{KonstanT21}. There are some additional aspects that requires particular attention. Firstly, the cyberspace where RSs are deployed is becoming increasingly complex and full of threats coming from various sources and of diverse nature, including cyber-attacks~\cite{de2023cybersecurity}, noisy and fake information~\cite{wang2022veracity} and system bias~\cite{chen2023bias}. This triggers the demand of \textit{trustworthy RSs (TRSs)} which aim to compete in the complex and challenging cyberspace. Secondly, stakeholders including users, owners and regulators of RSs, have increasingly higher demand for RSs \cite{AbdollahpouriAB20}. These stakeholders not only demand recommendation accuracy, but also need trustworthiness, including robustness, fairness, explainability, privacy-preservation, etc. Actually, trustworthiness is even more important than accuracy in some critical and sensitive domains, including finance and medicine, where highly reliable RSs are required. In practice, it has become a consensus both in the academia and industry that accuracy should
not be the only focus of an RS, and trustworthiness must be prioritised. These analyses have triggered the urgent demand of a new RS paradigm, i.e., TRSs. 

Some researchers have surveyed the existing work in the area of TRSs. For instance, Ge et al.~\cite{ge2022survey} systematically surveyed the techniques for key aspects related to trustworthy recommendation, including explainability, fairness, privacy-preserving, robustness and user controllability in recommendation. Similarly, Fan et al.~\cite{ge2022survey} conducted a comprehensive overview of
TRSs, specially focusing on six most important aspects including safety \& robustness, nondiscrimination \& fairness, explainability, privacy, environmental well-being, and accountability \& auditability. Jha et al.~\cite{jha2021survey} presented an exhaustive survey to emphasize the current research in the field on TRS models. Jin et al.~\cite{jin2023survey} surveyed existing methodologies and
practices of fairness in recommender systems, which is an important aspect of TRSs. These surveys mainly focus on summarizing existing work and progress in the filed of TRS, while lacking a high-level overview and a unified framework for TRSs, totally differentiating them from our work.     

In practice, researchers lack a systematic overview of TRSs, a rapidly developing field. There is neither existing work to systematically discuss the fundamental concepts, critical challenges, nor work to provide a unified framework for TRs, which are essentially quite important for the further development of TRSs. Therefore, it is in an urgent demand to provide a high-level overview on TRSs to sort out the fundamental concepts, key challenges and whole picture in this emerging area. To this end, this paper aims to provide a comprehensive overview of TRSs. To be specific, we discussed the motivation of TRSs in this section. Then, in Section 2, we will explore the fundamental concepts underlying TRSs, including the novel proposal of trustworthy recommendation ecosystem, aspects of TRSs, and examples of TRSs. In Section 3, we will present a systematic review of the paradigm shift in the field of recommendation, followed by an illustration of a four-stage perspective for TRS in Section 4. In Section 5, we will systematically analyse the critical challenges faced in building TRSs in each of the four stages. In Section 6, we will present a conceptual framework to build TRSs and then we will share some future directions in this vibrant area in Section 7. We will conclude this work in Section 8.

\section{The concept of trustworthy recommender systems \label{what}}
The term ``trustworthy'' describes an entity on which a subject can rely to be good, honest, sincere~\cite{trustworthy}. Specific aspects of ``trustworthy'' include: trustable, reliable, dependable, faithful, honourable, creditworthy, responsible~\cite{zhang2022trustworthy}. In general, a \textbf{TRS} refers to an RS on which its stakeholder can rely to be good, trustable, reliable, dependable, and faithful. 

A recommendation process is complex and interactive; it involves several entities, such as, users, items, the RS, and even the provider of items (e.g., sellers). All of them actually form a \textit{recommendation ecosystem} which enables the recommendation activities~\cite{AbdollahpouriAB20}. In practice, these parties also correspond to the necessary elements for recommendations: data (i.e., user/item/provider related information, e.g., user-item interactions), approaches and models (i.e., RS).

TRSs are actually the result of trustworthy RS approaches and models built in a trustworthy ecosystem. 
Although trustworthy RS approaches and models are pivotal for enabling trustworthy recommendations, they cannot survive without a trustworthy ecosystem. Most of studies in the literature only discussed trustworthy RS approaches while ignoring the significance of trustworthy ecosystem. For instance, Ge et al.~\cite{ge2022survey} provided a timely survey on the various techniques for enabling robust, fair, explainable or secure RSs while without touching trustworthy ecosystems or complex challenges in building TRSs, which greatly differentiates it from our work. To bridge this gap, we first propose a novel concept of trustworthy recommendation ecosystem, which is the foundation for building trustworthy RSs. In addition, trustworthy recommendation ecosystem can well differentiate trustworthy RSs from the generally discussed trustworthy artificial intelligence (AI) since it is derived from the core characteristics and properties of the recommendation scenario. Then, we illustrate the various aspects to specify the ``trustworthiness'' of RSs with an emphasis on the trustworthiness of recommendation approaches and models. Finally, we describe some specific and straightforward examples of trustworthy RSs in the real world.

\subsection{Trustworthy recommendation ecosystem.}
A recommendation process unfolds in a complex user-RS-item interaction scenario. Items are provided by various providers, including sellers, news source websites, and thus providers are an indispensable part of the process. As mentioned above, these four parties together form a basic ecosystem for recommendation activities and they interact with each other, possibly in a collaborative form, to complete the recommendation task. 
A trustworthy recommendation ecosystem includes trustworthy \textit{users, items, providers and RSs}, which will be illustrated successively in the following paragraphs.

\paragraph{Trustworthy users} Users consuming recommendations can possess diverse characteristics and perform various tasks in the cyberspace. Some users may not be trustworthy, and may try to obstacle the proper functioning of RSs. For instance, it is not uncommon to identify special users (e.g., internet water army and internet hacker) who may attempt to perform fake or false interactions with items to intentionally bias the recommendation models for a given personal and even malicious purpose (e.g., biased product promotions)~\cite{zhang2013commtrust}. 
Therefore, in order to build a trustworthy RS, only trustworthy users should be authorised to operate. Trustworthy users can be roughly characterized by two aspects: (1) the user profile information should be true, accurate, and users' behaviour on the recommendation platforms should reflect genuine preferences and intentions; (2) users should maintain a good reputation in their interaction behaviour~\cite{wang2008evaluating,wang2006trust}, e.g., they should write true and objective reviews or ratings for items, and pay for the purchased items on time.

\paragraph{Trustworthy items} Usually, there are massive quantities and diverse types of items on an online consumption platform. Items have diverse characteristics and can bring different impact to the RS stakeholders, which are individual users, the local community and even the whole society. Although most of the items can well satisfy users' needs and can have a positive impact on the stakeholders interests in the long run, some items may produce a negative impact on either the individual users, the community or the society. For example, additive electronic games usually match young children's preferences well. But these games often create addiction and thus can negatively affect users' lives, work or studies, hence resulting in significant negative impact on users and their family. Another example is fake news, which may match some minority users' stance and reading preferences. But they mislead the public, and may cause serious social problems, and should not be recommended to end users~\cite{wang2022veracity}. To this end, only trustworthy items should be recommended by a trustworthy RS. Trustworthy items can be characterized by two aspects: (1) the item-related information (e.g., item attributes) used for recommendations should be true and precise, so that the intrinsic characteristics of all items can be easily captured; 
(2) the candidate items to be recommended to end users should be trustable and responsible, namely the items must not bring potential negative impacts to the stakeholders of RSs.

\paragraph{Trustworthy providers} Providers are the party who provide items in online platforms. For example, a provider may be a seller in an e-Commerce platform, or a news source website which releases original news in the news domain. In the real world, some providers are not trustworthy since they may provide fake or poor-quality items, leading to negative impact to the end users and the society. For instance, some sellers may sell fake products while some news websites may release fake news. Obviously, trustworthy providers are necessary for building trustworthy RSs and they can be characterised by two aspects: (1) the providers should be responsible for the quality and potential impact of the items they provide. For instance, the sellers should provide products or services with guaranteed quality, while the news websites should provide verified true news only while combating fake news; (2) it is important for providers to maintain a good reputation through their historical transaction or releasing behaviours~\cite{zhang2015reputationpro,wang2015social}. 

\paragraph{Trustworthy RSs} Recommendation approaches which are usually built on data mining or machine learning models should be reliable and stable. Specifically, they should not only be able to accurately capture users' preferences and items' characteristics so as to provide accurate recommendation services to end users, but also perform stably all the time, even in the most complex, dynamic and challenging context (e.g., the cyberspace facing frequent cyber-attacks). Trustworthy approaches and models are the most important part in trustworthy RSs and they can be characterised by multiple specific aspects, which will be discussed in detail in the next part.  

\subsection{Aspects of trustworthy recommender systems.}\label{aspect} 

In some related areas, such as, trustworthy AI, many different aspects have been proposed to specify ``trustworthy'' from perspectives including regulations, mechanism, and technology~\cite{zhang2022trustworthy}. Several aspects related to the technology perspective have been commonly recognized, including robustness, fairness, explainability, transparency, privacy, accountability, and responsibility. Since RSs are a specific application area of AI, the aspects used to define trustworthy AI are also applicable to characterise trustworthy RSs. However, due to the very special and specific data characteristics, work mechanism and computation task of an RS, there are additional aspects, such as, human's perception and trustworthy evaluation, which are specific to the RS scenario and can well distinguish trustworthy RSs from trustworthy AI. Next, we illustrate the various aspects for characterizing trustworthy RSs.

\begin{itemize}

\item \textbf{Robustness} indicates an RS that has strong fault-tolerant capability to survive attacks, noisy and fake information (e.g., such as users' arbitrary behaviours and fake reviews/ratings on items)~\cite{mobasher2007toward,liu2022federated}, which often commonly affect the input data used for recommendations.

\item \textbf{Fairness} indicates that the RS is able to reduce or remove possible bias towards certain groups of stakeholders (e.g., the disparities in treating user with different demographics or items with low and high popularity) which may be present in the input data, the recommendation model or in the recommendation results~\cite{mehrabi2021survey,ge2022explainable,wang2023survey}. Consequently, the final recommendation results can be assessed as fair from both the user side and the provider side.

\item \textbf{Transparency} generally means that the functioning mechanism and the recommendation model are transparent to all the stakeholders of the RS, and it is not perceived as a ``black box''~\cite{balog2019transparent,di2022recommender}. Transparency can greatly reduce the potential risks of deploying ineffective RSs in real-world domains, especially in some sensitive and critical domains like finance and healthcare domains. 

\item \textbf{Explainability} means that the recommendation mechanism, recommendation models and the recommendation results are well-explained so that the stakeholders of the RS can properly understand how and why the recommendations are generated~\cite{zhang2020explainable,li2023personalized}. 

\item \textbf{Privacy and security} means that the RS is able to effectively protect the personal privacy of relevant stakeholders~\cite{shin2018privacy,imran2023refrs}. This aspect is even more important in RS, compared to other web applications, since RSs are consuming a large amount of profile and behaviour, which often contains a sensible users' characteristics.

\item \textbf{Responsibility}. The development and deployment of RSs should be conducted in a responsible manner (e.g., follow the regulations and law), so that the RS will not explicitly or implicitly harm anybody. Moreover, the generated recommendation results should be beneficial to the all stakeholders~\cite{elahi2021towards}. For instance, fake news should not be recommended to any users even if it may interest some ones, or addictive games should not be recommended to young children either. 

\item \textbf{Human's perception of trustworthiness}, relates to the subjective perception, which the relevant stakeholders have, of the trustworthiness of an RS. This is unique to RSs and is of great significance. In fact, an RS is actually involving a human-machine interaction~\cite{jannach2016recommender}, and whether an RS is trustworthy or not, ultimately depends on human's feeling and perception. However, almost all of the existing approaches to model trustworthy RSs have been developed by assuming a single point of view, that of the machine~\cite{wu2012hysad}, i.e., the RS models and algorithms, while have ignored the human's factor. Moreover, existing work often mechanically computes a numerical measurement value to quantify the trustworthiness of RSs. Such lack of consideration of the human's factor triggers the urgent demand 
to take human's perception into account when talking about trustworthy RSs. 
\begin{figure*}
	\centering
	\includegraphics[width=8.8cm]{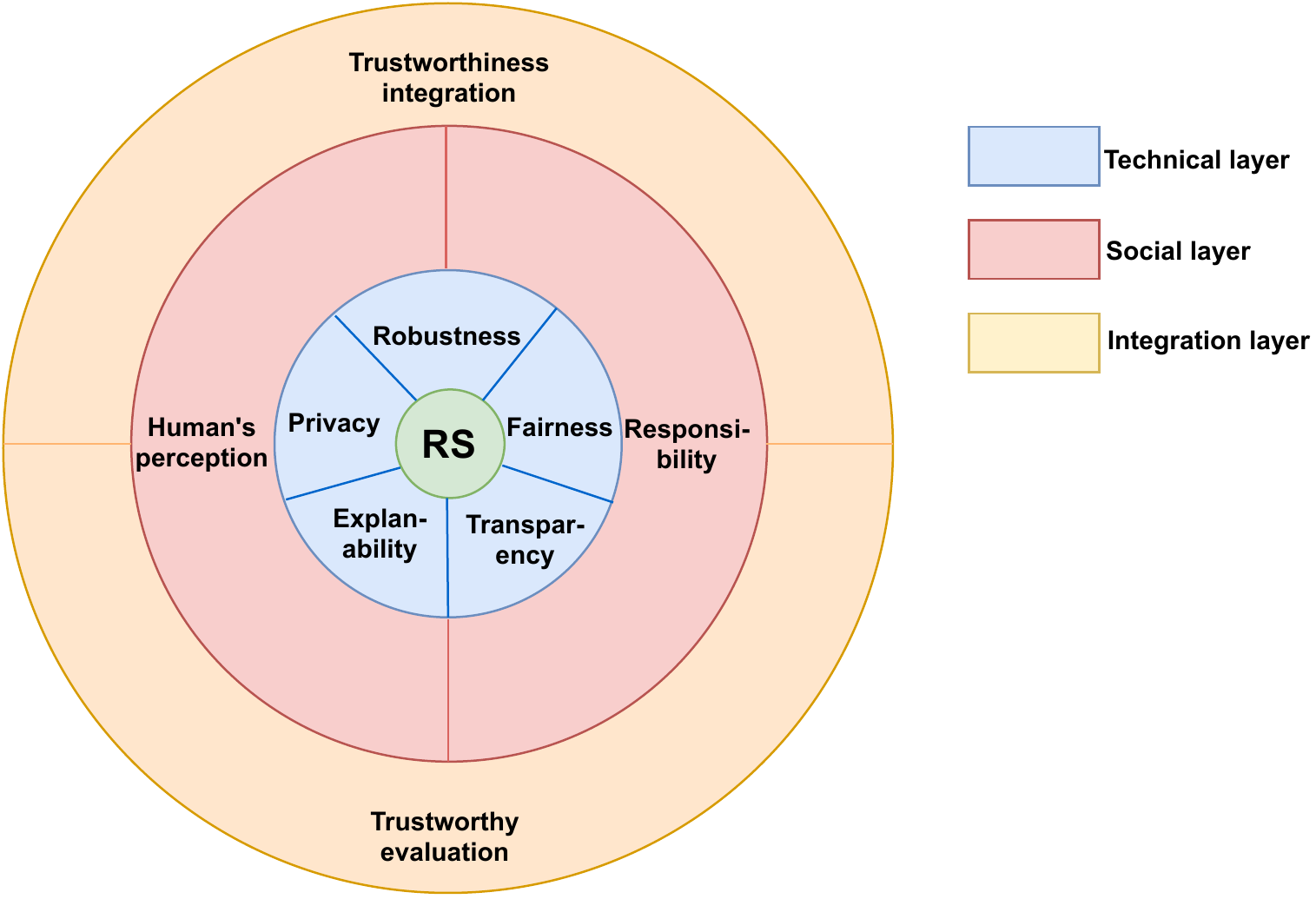}
	\vspace{-1em}
	\caption{The various aspects to characterise TRSs and the relations among them. According to the characterization perspective, they can be organized into three layers. }\label{trs_Conc}
	\vspace{-1.5em}
\end{figure*}

\item \textbf{Trustworthiness integration.}
Although various aspects and perspectives have been proposed to characterise trustworthy RSs, they are mostly separated from each other. 
In addition, most of the existing studies on trustworthy RSs have focused on one or two aspects of trustworthiness alone. For instance, fairness-aware RSs and explainable RSs only focus on the fairness and explainability aspect respectively. However, accounting for only one or two aspects can not lead to a truly trustworthy RSs. In order to build truly trustworthy RSs, all the aforementioned different aspects, including both objective and subjective aspect, should be integrated together organically and effectively towards to a unified trustworthy RS framework.

\item \textbf{Trustworthy evaluations.} On one hand, both the evaluation methods and the evaluation metrics must be reliable to well and precisely reflect the performance of the RS. One example is that most offline evaluations are performed under a very ideal situation and thus they cannot actually indicate how an RS will perform in the real-world online scenarios~\cite{GunawardanaS15}. On the other hand, new evaluation protocols and metrics are in demand to evaluate the trustworthiness of an RS so that all the aforementioned aspects can be appropriately validated. 

\end{itemize}

Although these different aspects touch a variety of issues and often correspond to different elements and stages of an RS, they are closely inter-connected and must be jointly addressed to contribute to the ``trustworthiness'' of the RS, as illustrated in Fig. ~\ref{trs_Conc}. For instance, fairness often relates to explainability. In fact, the fairness perspective (e.g., users' gender or item popularity) that  is considered and how fairness is assured should be explainable to the end users, so that they can clearly understand and accept the generated fair recommendation results.

\subsection{Examples of trustworthy recommender systems in the real world.}

For decades, RSs have been widely exploited in nearly all information systems that we use daily for work, study or entertainment. In recent years, along with the occurrence of an increasing number of cyber threats including cyber attacks, fake or false online information, bias in the cyber space, trustworthy RSs are an evident need. There is a variety of examples of real-world trustworthy RSs across different application domains. In E-commerce domain, one typical example is the robust RS which can tame fake ratings and reviews~\cite{lyu2021reliable} in platforms like Ebay or Amazon. This type of RSs detects fake ratings and reviews inserted in the platform and discard them when generating trustworthy recommendations. In media domain, a typical example is a recently proposed fact check technology (e.g., Google fact check tools) for the news domain, which introduces a series of techniques to check the veracity of news on the web so that only verified true news will be recommended to end users~\cite{giansiracusa2021tools}. 
Another typical example is a veracity-aware news recommender system published at the 2022 Web Conference~\cite{wang2022veracity}. This system checks the veracity of each news piece before generating recommendations, so that only checked true news will be recommended to end users. 




\section{The Recommendation Paradigm Shift} 
\subsection{From accuracy-oriented RSs to trustworthy RSs}

RS research can date back to 1990s when the first collaborative system ``Tapestry'' was reported in the literature~\cite{goldberg1992using}. From early 1990s to the middle of 2000s, several algorithms have been developed for enabling content-based or collaborative filtering recommendations~\cite{2022rsh,wu2022survey}. From 2008 to 2016, driven by the popular Netflix challenge, matrix factorization methods have dominated the scenario for a long time. Since the middle of 2010s, benefiting from the rapid development of deep learning techniques, deep learning based RSs have become the main stream in recommendation domain~\cite{zhang2019deep}.

Along with that technological shift and revolution of RSs, the focused research problems, challenges and evaluation mechanisms of RSs have also changed. Since the first studies on RSs to early 2010s, the accuracy of recommendation results has been the most important and often the only evaluation criteria that was used to evaluate an RS. Specifically, that means that the researchers tried to assess whether the recommended items were truly preferred (e.g., clicked, purchased, highly rated) by users. 
During this period, nearly all efforts were devoted to develop more accurate RS models and algorithms, such as more accurate matrix factorization algorithms to predict the unknown user-item ratings with a smaller and smaller prediction error. Then, starting from early 2010s, more and more researchers realized that the accuracy criterion is insufficient for evaluating an RS, and thus new evaluation dimensions, including diversity and novelty were proposed~\cite{adomavicius2011improving,GunawardanaS15}. As a result, the evaluation of an RS moved from single-criteria evaluations, based on accuracy, to multi-criteria evaluations based on a range of dimensions, such as accuracy, diversity and novelty, while accuracy still was considered as the most important criteria.  

Since late 2010s, due to an increasing level of threats in the cyberspace, coming from cyber attacks, noises, and biases, the trustworthiness of RSs have attracted increasing attention from both academia and industry. As a result, a growing number of studies have focused on trustworthy RSs, from various perspectives, such as robustness, fairness, explainability or privacy. For example, some work aimed to build robust RSs which can fight against shilling attacks~\cite{gunes2014shilling}, and fairness-aware RSs to generate fair recommendations for users with different demographics~\cite{xiao2017fairness}. Some other work have instead focused on building explainable RSs to generate user-understandable recommendation results~\cite{zhang2020explainable}, or secure and privacy-preserved RSs which puts a special concern on users' privacy~\cite{shin2018privacy}. Obviously, the research focus and the ultimate goal of these studies is no longer centered on the improvement of recommendation accuracy. Instead, they have changed the direction towards enhancing the trustworthiness of RSs. This has actually revolutionized the research focus and the evaluation perspectives, from  accuracy-oriented recommendations towards  trustworthiness-oriented recommendations, leading to a recommendation paradigm shift. 

In practice, the trustworthiness of RSs has been attracting much attention from different sectors including academia, industry and government. A typical example is that the Chinese government released regulations on recommendation algorithms for internet information services in early 2022~\cite{TRS_law}. A large proportion of the regulations emphasised the significance of trustworthy RSs to be deployed in the real-world business applications.

\subsection{From trust-aware recommender systems to trustworthy recommender systems}
 
Trust-aware recommender systems mainly consider and leverage the trust relationships between humans and organizations to enhance the recommendation performance. Here the trust relationships can vary, some typical examples include the trust relations between different users on online or offline social networks (e.g., friends on facebook, or colleagues in one organization), the trust relationships between users and comments, ratings on items from others, the trust relationships between customers and sellers~\cite{zhang2012trust,zheng2014trust}. Since early 2000s, a variety of studies~\cite{massa2007trust,dong2022survey,ahmadian2022reliable,zhang2015reputationpro} have been done to investigate how to well to model and leverage such diverse trust relationships to improve the performance of RSs.     

Differently from trust-aware RSs, trustworthy RSs aim to build reliable RSs which can be accepted by humans, which were proposed in late 2000s for the first time~\cite{mobasher2007toward}. However, early definitions on trustworthy RSs mainly focused on one single aspect of trustworthiness, namely robustness~\cite{mobasher2007toward,wu2012hysad}. So trustworthy RSs in the early stage mainly refers to those RSs which are robust while facing various types of attacks in the cyberspace. However, they are not fully trustworthy since there are some other important aspects contributing to trustworthiness that were initially overlooked, including the aforementioned fairness, explainability, privacy and responsibility.            

Therefore, it is of great theoretical and practical significance to develop the next-generation trustworthy RSs which go beyond the existing robustness-oriented ones. To this end, in this paper, we propose novel trustworthy RSs in which all the nine aforementioned aspects contributing to trustworthy RSs are considered, leading to a significant step towards the implementation of truly trustworthy RSs.

\section{Trustworthy Recommender Systems: A Four-stage Perspective}\label{Four-stage}

\paragraph{Classical Recommendation Process} Generally speaking, a typical machine learning-based recommendation process comprises the following four successive and inter-connected stages:
\begin{itemize}

\item \textbf{Data preparation stage,} including data collection and preprocessing; 

\item \textbf{Data representation stage,} which often learns an informative representation of the raw input data as the input of the downstream recommendation model; 

\item \textbf{Recommendation generation stage,} which deploys some data mining or machine learning based prediction models to predict the unknown user-item interactions, e.g., ratings, clicks; 

\item \textbf{Performance evaluation stage,} which employs various evaluation methods, including online and offline evaluations, to evaluate the recommendation results from different perspectives such as accuracy, diversity, and fairness. 

\end{itemize}

\paragraph{Trustworthy Recommender Systems}

To be trustworthy, an RS should be trustworthy in each of the aforementioned four stages. Each stage has its unique computation task, characters and goal, so the ``trustworthiness'' of different stages often has different specific meanings and focuses on different aspects of the aforementioned nine aspects (cf. Section~\ref{aspect}). For instance, data preparation stage often focuses more on the ``robustness'' aspect, since the data preparation methods should be robust enough to survive to noisy and fake data, possible coming from attacks. While the recommendation generation stage should focus more on aspects like ``fairness'', ``transparency'' and ``robustness'', since fair, transparent and robust RS models and algorithms are required in this stage for generating recommendations. Next, we try to characterize a trustworthy RS from a four-stage perspective while the specific meaning of ``trustworthy'' in each stage will be highlighted. 

\begin{itemize}
\item \textbf{Robust and secure data preparation}, on one hand, is necessary to combat  noisy and fake information, and bias in the raw data, generated in the cyberspace, and survive to possible attacks on the data. In this way, the prepared data for recommendations can accurately reflect users' preferences and items' characteristics. On the other hand, the users' data should be managed and processed in a secure way so that  sensitive and privacy information will not be leaked.    

\item \textbf{Robust and explainable data representation,} which employs robust and explainable data representation models to precisely learn informative and interpretable latent representations of the original input data in a stable way regardless of the potential attacks. In most cases, the learned latent representations are latent vectors without an explicit semantic meaning, which then negatively impacts on explainability. In addition, it is often not very clear which information is encoded in the representations. Therefore, the ``explainability'' aspect should be emphasized in this stage.     

\item \textbf{Fair, transparent, explainable and responsible recommendation generation}, in which fair, transparent and explainable RS models should be adopted to provide fair exposure opportunities of all items to different users in a transparent, explainable and responsible way. In addition, the models should also be able to provide straightforward explanations by using texts or images corresponding to the recommendation results, to faithfully explain how the recommendations are generated and why they are appropriate for a given user. More importantly, only those items with positive impact to stakeholders can be recommended. 

\item \textbf{Trustworthy evaluations} comprise two-sided evaluations: (1) \textit{technical evaluation}, and (2) \textit{ethical evaluation}. The technical evaluation aims to evaluate the recommendation performance from some technical perspectives such as accuracy, diversity, novelty, and explainability. More trustworthy and reliable evaluation protocols and metrics are required to evaluate such technical performance. In contrast, the ethical evaluation aims to evaluate recommendations from the ethical perspective such as the responsibility and social impact. For instance, whether the recommendation activities and the recommendation results are responsible and beneficial to the relevant stakeholders. Towards trustworthy RSs, ethical evaluations are necessary and of great significance to the sustainable development of the community and the society. Some examples of unethical but maybe preferable recommendations include the recommendations of fake news~\cite{wang2022veracity} and the recommendation of addictive games to young children. To the best of our knowledge, this is the first time the proposal of an ethical evaluation of RSs is put forward in the literature.   
\end{itemize}

\section{Challenges for Building Trustworthy Recommender Systems}
\begin{figure*}
	\centering
	\includegraphics[width=13.8cm]{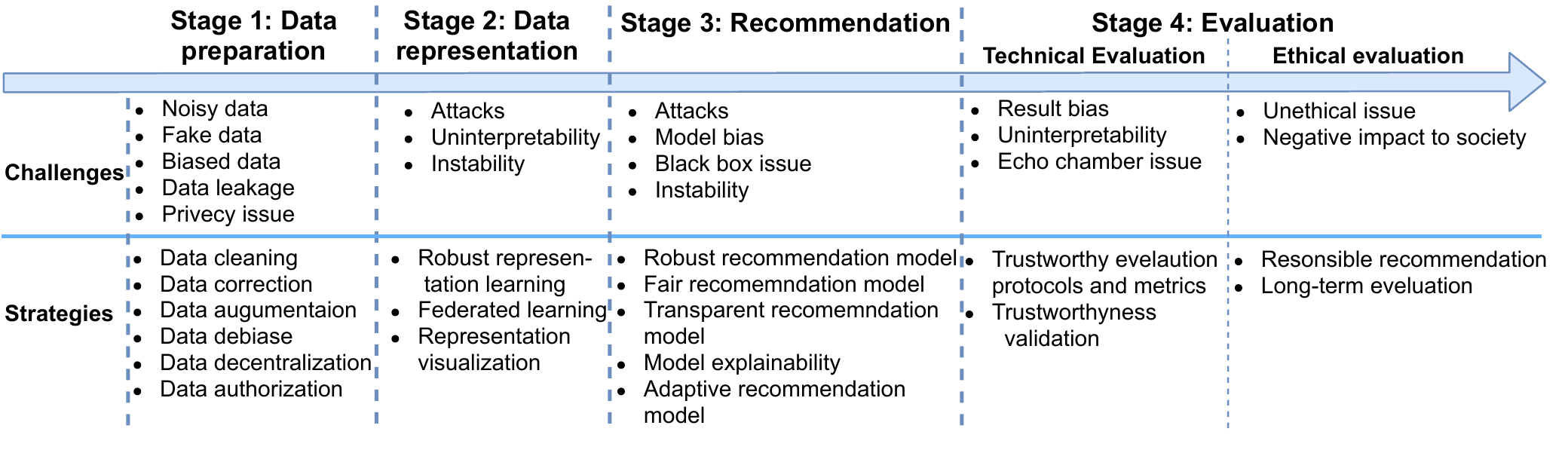}
	\vspace{-1em}
	\caption{The challenges and corresponding strategies for building TRSs.}\label{trs_ch}
	\vspace{-1.5em}
\end{figure*}
Although the significance and urgency of trustworthy RSs have been commonly recognised, the research in this area is still in an early stage. A variety of challenges and research problems remain to be properly addressed in each of the aforementioned four stages. We therefore analyze 
the particular challenges of each stage.

\paragraph{Challenges in data preparation stage.}
Before building an RS model, a necessary step is to collect and prepare data to train and test the model~\cite{wang2023data}. Such data often contains user-item interactions such as users' clicks, purchases, ratings, or comments on items, and side information such as users' attributes, social relations, and items'
features~\cite{JannachPRZ21}. 
The data is often collected from online consumption platforms such as amazon.com, youtube.com, and abcnews.go.com. Due to the large amount of diverse users (e.g., 10M+ users) on each platform, it is inevitable that some users' behaviours may be noisy and unreliable (e.g., randomly click or view a set of items on amazon.com), or there may be some malicious users (e.g., internet water army) who have performed malicious behaviours, e.g., posting fake ratings or comments to some targeted items or sellers. These factors bring noise, bias, and fake information into the interactions or side information, making the data used for recommendations unclean and unreliable. As a result, the data can not accurately reveal the users' preferences and item characteristics, and thus misleads the downstream recommendation models.

Therefore, a big challenge at the data preparation stage is to effectively detect and remove critical data, e.g., noisy, fake, or biased data, and retain the genuine data. This task is quite challenging since this critical data usually accounts for a quite small proportion of the whole population, making it very difficult to discover it, especially when critical data looks quite similar to the genuine data. Possible strategies to address this challenge include a set of conventional and emerging data preprocessing techniques. 
For instance, some advanced data augmentation and data debias methods~\cite{SchnabelSSCJ16}, including counterfactual data augmentation~\cite{wang2021counterfactual} and contrastive learning~\cite{tian2020makes}, have been extensively studied to remove the bias in raw data in recent years. 


Another challenge at this stage is related to the data privacy and security issue. This is especially important in real-world RS applications since the data used for recommendations often contains a large amount of users' private information which should be protected. To be specific, on one hand, during the collection of raw data, it is important to ensure that only the really required data will be collected. On the other hand, when a large amount of data is collected, it is important to build a safe and strong infrastructure and mechanism to properly store and manage it, including secure and powerful database, and rigorous data access mechanism. Another way to address the data privacy and security issue is to introduce federated learning~\cite{yang2019federated} and edge computing~\cite{shi2016edge}. In this way, the data used for recommendations can be decentralized to various edge devices such as users' mobile phones where the individual users' privacy can be well protected. 

\paragraph{Challenges in data representation learning stage.}
Data representation aims to learn an informative numerical vector in a latent space to well represent each data point in the original input dataset. For instance, the well-known representation learning model Word2Vec~\cite{rong2014word2vec} learns a latent numerical low-dimensional vector to represent each word.      

In advanced machine learning, especially in deep learning area, data representation learning is a prior step of all the downstream learning models and tasks. The quality of data representation directly determines the performance of the learning models. Therefore, machine learning based RS models often consists of two main steps: data representation learning and prediction. The output of the first step is taken as input of the second for the prediction task. 
For instance, in matrix factorization or neural network based RS models, each user/item is represented by a K-dimensional latent vector which is learned by mapping the user/item ID into a latent space~\cite{NingDK15}. The main challenges in representation learning stage lie in the following three aspects.

First, how to learn accurate and informative representations to precisely retain the original information in the raw data? For instance, the similarities/distances between the raw data points should be accurately reflected by their corresponding representations in the latent space. Towards this challenge, some advanced representation learning techniques including contrastive representation learning~\cite{xiong2020loco} are able to learn more accurate and informative representations.  

Second, how to learn interpretable representations? One common and critical issue with latent representations is the lack of explainability. It is often not clear what the semantic meaning of the latent vectors is, making the learned representations uninterpretable for humans. Sometimes, it is not clear what kind of information is encoded in the latent vectors either, which brings some potential risks for the real-world applications of RS models. To address this challenge, some explainable representation learning models and techniques including data visualization can be developed to disclose the patterns encoded in the latent representations. 

Third, how to combat attacks on representation learning models? In general, nearly all machine learning models including representation learning models may suffer from some external attacks, especially  those deployed in real-world applications. The attackers may quickly generate a large amount of fake data to intentionally mislead the models and thus downgrade the quality of the learned representations. 
Some possible strategies include robust representation learning techniques, such as adversarial representation learning which fights against attacks in an adversarial way~\cite{donahue2019large}.

\paragraph{Challenges in recommendation generation stage.}
In this stage, the data representations learned in the previous stage will be imported into the selected RS model for generating recommendations. The key challenge lies in how to build a \textit{robust, fair, transparent, explainable} and \textit{responsible} RS model which can work in a trustworthy way in a complex and dynamic environment.

First, similarly to what was observed for the representation learning stage, RS models also easily suffer from various attacks. So the first challenge is how to build robust RS models which can survive various attacks especially those malicious attacks and adversarial attacks in the complex cyberspace where RSs are deployed. 

Second, some particular design and work mechanisms of the RS models may lead to biased RS models and thus generate biased recommendations. Therefore, another challenge in this stage lies in how to effectively remove the bias. Various model debias techniques like auto-debias which learns the debias parameter from the data~\cite{Chen2021Auto} can address this issue to some degree.

Third, advanced machine learning based RS models, especially deep learning based ones often run in a black-box mode, and thus the computations in the model are not transparent and explainable to humans. This makes it hard to understand how an RS model is working and what potential risks and disadvantages the model may have. 
Hence, this may bring potential risks for the applications of RS models in real-world, especially in some social and economic domains which are critical, such as, healthcare and finance, where a small mistake may lead to a large loss (e.g., the loss of a life or a large sum of money). Hence, the third challenge in this stage lies in how to build transparent and explainable RS models whose work mechanism are well understandable to humans. In this way, how and why the recommendations are generated can be well disclosed and visible to end users, which greatly reduces the risks of the employment of RSs in real-world business.

Fourth, considering the real-world application context, some recommendation behaviours may be profit-driven while ignoring the ethical or even the legal issues (e.g., recommending additive games to children). Some recommendation results may be accurate but not responsible, such as recommendations of fake news to particular users~\cite{wang2022veracity}. Hence, how to ensure responsible recommendation behaviours and recommendation results which are for social good is another big challenge in this stage.     

\paragraph{Challenges in the evaluation stage.}
In this stage, the recommendation results, e.g., predicted user-item rating values, or a list of ranked items, generated by an RS model are evaluated to measure its performance. Conventionally, the evaluation metrics could be a variety of quantity metrics of concern such as accuracy, and diversity, which can directly measure the performance from the technical perspective. However, as mentioned in the Introduction, trustworthy evaluation should go beyond  conventional technical performance evaluation. Instead, they should contain both \textit{technical evaluation} and \textit{ethical evaluation}.

For the technical evaluation, existing evaluation protocols and metrics for RSs are mostly accuracy-oriented~\cite{wu2022survey}. They are not sufficient to comprehensively evaluate an RS, especially in the complex cyberspace which often contains noises,
attacks, and bias~\cite{huang2018systematically}. In such a complex context, the stakeholders of an RS usually focus  not only on accuracy, but also on trustworthiness. 
Hence, the effective evaluation of the trustworthiness of an RS has a great social and economic significance. However, the evaluation of the trustworthiness of an RS is quite challenging, and involves multiple aspects, such as the robustness, fairness, explainability and privacy. Some of them are hard to be quantified in a traditional online or offline experiment. In addition, different application scenarios often have different requirements on the trustworthiness of the RS and focus on different aspects. For instance, in an online e-commerce platform where attacks and noise behaviours are routinely detected, the robustness of the RS should be a primary attention point. Conversely, in an offline RS for medicine treatment, the transparency of the RS model and the explainability of the recommendation results may be more important. Therefore, it is of great importance to define novel trustworthiness evaluation schemes for RSs, including both new evaluation protocols and evaluation metrics for comprehensively and systematically measuring the trustworthiness of RSs.

Ethical evaluation aims to evaluate an RS from the ethical and social perspectives. Specifically, they measure whether the recommendation behaviours and recommendation results will have a positive impact on the various stakeholders. 
Here, the impact of an RS could be on the recognition or the behaviours of end users, and it could be explicit or implicit, short-term or long-term. Ethical evaluations are quite challenging since the various impact and influence of RSs on both individual level and the society level are very difficult to accurately capture and quantify~\cite{HazratiR22}. New and well-designed ethical evaluation methods are in demand for RSs, which may combine both qualitative and quantitative evaluations.

\section{A Framework for Trustworthy Recommender Systems}

To build a trustworthy RS is therefore a challenging task. Not only various aspects of trustworthiness, such as robustness, fairness and security, and the complex relations among them, should be well considered in a unified way, but also new and trustworthy evaluation protocols and metrics are in order. Each aspect often involves specific models and techniques. For instance, robustness is often achieved via adversarial learning techniques such as generative adversarial networks~\cite{DeldjooNM21}. Due to its extreme complexity, it is challenging to design a very specific model for trustworthy RSs which can well handle all the aspects. 
However, a systematic methodology to build a comprehensive trustworthy RS while considering all the different aspects in different stages is in a compelling challenge and there is the need of a unified framework for trustworthy RSs. Hence, to bridge this gap, we propose a conceptual framework to support trustworthy RSs. 

\begin{figure*}
	\centering
	\includegraphics[width=13.8cm]{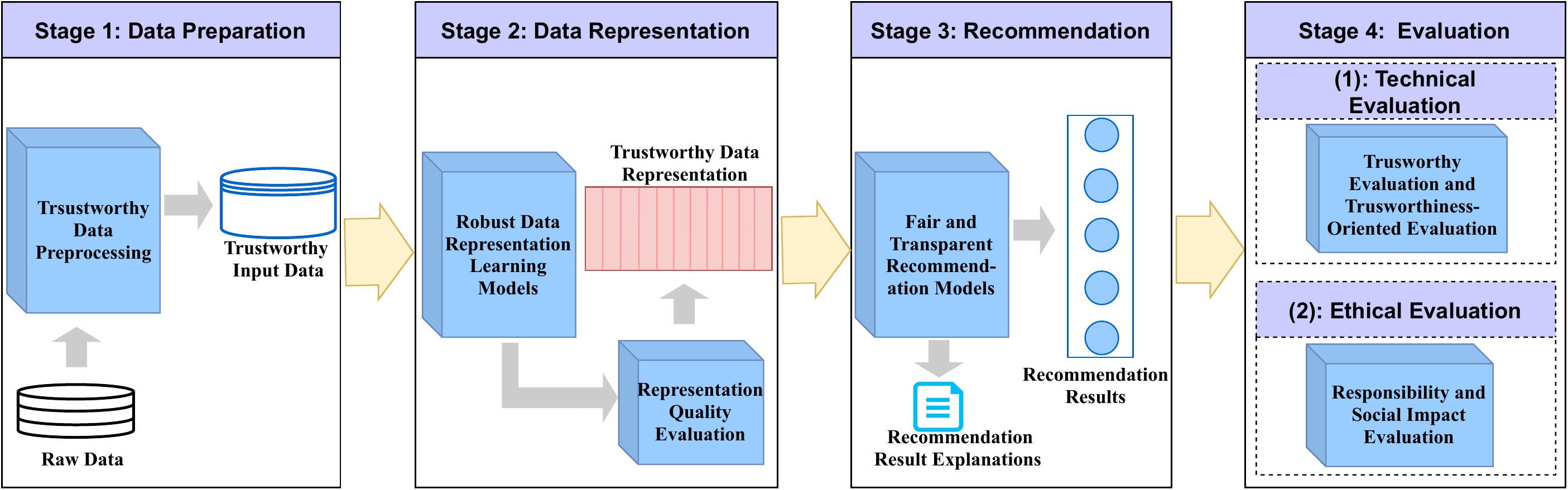}
	\vspace{-1em}
	\caption{A conceptual framework for building TRSs.}\label{trs_framwork}
	\vspace{-1.5em}
\end{figure*}

Consistent with the four stages of recommendation process discussed in Section~\ref{Four-stage}, our proposed trustworthy RS framework consists of four stages: (1) \textit{trustworthy data preparation}, (2) \textit{robust and explainable data representation}, (3) \textit{fair, transparent, explainable and responsible recommendation generation}, and (4) \textit{trustworthy evaluation}, as is is depicted in Figure~\ref{trs_framwork}. 

\paragraph{Trustworthy data preparation.}
Trustworthy data preparation aims to first collect raw data (e.g., user-item interaction data and side information data) from online platforms in a trustworthy way (e.g., privacy-preserved) and then transfer it to trustworthy input data (e.g., clean, unbiased and reliable data) for the downstream recommendation tasks. Note that the collected raw data often contains noises, bias and false information, etc. Some conventional and advanced data preprocessing techniques can be employed to support trustworthy data preparation. For instance, some data cleaning methods including pattern-based error detection~\cite{he2005fp} and machine learning based outlier detection~\cite{pang2021deep} can be utilized to detect and remove those noisy and fake data in the raw dataset. In recent years, some advanced machine learning approaches like counterfactual data augmentation~\cite{wang2021counterfactual} and contrastive learning based data augmentation~\cite{tian2020makes} have been commonly utilized to reduce the bias in input data for recommendations. 

\paragraph{Robust and explainable data representation.}
In principle, there are two specific steps for achieving trustworthy data representations: (1) representation learning to learn user and item representations, and (2) representation quality evaluation to evaluate the learned representations. In RS area, 
the evaluation step is often incorporated into the downstream recommendation task, i.e., the representation quality is measured by the quality of recommendation results.  


For the representation learning step, as discussed in Section~\ref{Four-stage},  trustworthy can be  specified into robust and explainable. On the one hand, to support the robustness of data representation learning models so that they can perform stably when facing attacks and noises, one typical solution is adversarial representation learning models~\cite{donahue2019large}. In addition, to remove the possible biased and noisy information hidden in the representations, some representative debiasing and denoising models including Denoising Autoencoder (DAE)~\cite{lu2013speech} can be deployed for more accurate representation learning for recommendations. On the other hand, in order to make the learned latent representations more interpretable for humans, some data visualization techniques such as t-SNE~\cite{van2008visualizing} can be helpful to explicitly reveal the hidden patterns in latent data.

\paragraph{Fair, transparent, explainable and responsible recommendation generation.}

In the recommendation generation stage, the core task is to design and develop trustworthy RS models. As discussed in Section~\ref{Four-stage}, trustworthy in this stage can be characterised along four dimensions: fair, transparent, explainable, and responsible. We illustrate them in the following.

First, a trustworthy RS model must be free of bias so that it can perform equally well for different users and items. 
In support of this, disentanglement learning based approaches~\cite{ma2019learning} can be used for removing bias in RS models. In addition to RS models, bias may also exist in recommendation results, which could be reduced by re-ranking techniques based on some particular constraints.


Second, trustworthy RS models should be transparent. 
To support the transparency, some set-based techniques can be employed to explicitly present the RS models to end users via natural language~\cite{balog2019transparent}. Building transparent RS models is quite challenging and more studies are deserved. 

Third, to support the explainability of recommendation results, the output of recommendation generation stage should be two-sided: (1) the normal recommendation results, e.g., a list of selected candidate items, and (2) corresponding explanations of the recommendation results to well explain how and why the recommendations are generated. The explanations should be straightforward and well-understandable, either in textual or visual formats~\cite{lu2019graph}. 

Last but not the least, to be trustworthy, the recommendation generation process must be responsible in terms of both the recommendation behaviours and the recommendation results. On one hand, some additional constraints and criteria can be employed to ensure the above aspects towards social good. For instance, some models can be developed to filter out those fake products, additive games and fake news from recommendation lists. On the other hand, some regulations may be enforced to guide the harmonic development and deployment of RS techniques.

\paragraph{Trustworthy evaluation.}
Evaluation is an important step for ensuring the quality of recommendation results generated by RS models. As depicted in Figure~\ref{trs_framwork} and discussed in Section~\ref{Four-stage}, both technical evaluation and ethical evaluation are required for a trustworthy RS. Regarding technical evaluation, most of the existing evaluations are accuracy-oriented, while trustworthiness evaluations are in their early stage. Although a few metrics have been defined to measure one specific aspect of trustworthiness, such as fairness~\cite{leonhardt2018user}, there is no unified method to systematically evaluate multiple aspects as discussed in Section~\ref{what}. New evaluation protocols and metrics are in demand for comprehensive trustworthiness evaluations.   


Regarding ethical evaluation, to the best of our knowledge, there is no existing work reported in the literature. Ethical evaluation mainly evaluates the social impact of recommendation behaviours and results. For instance, whether the recommendation behaviours are legal and ethical or if the recommendation results lead to some good/bad impact (e.g., lead users to become addicted to video games). Ethical evaluation is full of complexity and dynamics, in which the users' cognition, perception and behaviours should be taken into account. To this end, some possible solutions include user studies, and the combination of qualitative and quantitative studies.

\section{Future directions}

\textit{Truly trustworthy recommender systems.} Although there are existing work aiming at building trustworthy RSs, nearly all of them focus on only one single aspect (e.g., fairness or explainability). For instance, some early work on trustworthy RSs only focuses on the robustness of RSs~\cite{mobasher2007toward}, aiming to build resilient RSs to combat attacks. More recently, some other work focuses on either explainability~\cite{zhang2020explainable}, security~\cite{lam2006you} or fairness~\cite{leonhardt2018user}. However, to be truly trustworthy, usually all these various aspects (cf. \ref{what}) of trustworthiness should be carefully considered in a unified way. Therefore, more analyses are in demand to build truly trustworthy RSs.

\textit{Human-centered trustworthy recommender systems.}
Although some studies have attempted to build trustworthy RSs, almost all of them only mechanically developed the so-called trustworthy models from the machine perspective, while ignoring the human factor. However, weather an RS is trustworthy or not should be ultimately determined by the relevant stakeholders, rather than by the RS model itself. Therefore, human's perception and judgement should be well incorporated when developing and evaluating a trustworthy RS, which deserves a deeper investigation.   

\textit{Trustworthy recommendation ecosystems.} As discussed in Section~\ref{what}, trustworthy RSs can not survive without a trustworthy recommendation ecosystem. However, almost all the existing work on trustworthy RSs only focus on the development of trustworthy RS models while ignoring other important elements, e.g., trustworthy items and providers, in the ecosystem. Therefore, more studies in the development of trustworthy recommendation ecosystems are in order.


\textit{Multi-granular fairness aware trustworthy recommender systems.}
Fairness is essentially multigranular and exists at different levels~\cite{mehrabi2021survey}, such as the high-level fairness between different groups, and low-level fairness between different individuals within each group. A really fair RS should be fair in all the different granularities rathern than only a single granularity as done in most of the existing work.  
Hence, to comprehensively model multigranular fairness is of great significance for building truly fair RSs.

\textit{Fine-grained personalized recommender systems.}
Although it is a common sense that personalization is the core aspect and the basis of RSs, the recomemndation results of most existing RSs are not personalized enough since they are mostly preferred by the majority users only while unsatisfied by the minority ones. Therefore, it is in highly demand for designing innovative RS models and evaluation approaches to ensure fine-grained personalization of recommendation results which can satisfy nearly each user. 


\textit{Responsible recommender systems.} Most of existing RSs are technique-driven, which are built on advanced machine learning techniques and are evaluated from the technical perspective only. The social impact of RSs are hardly considered during the design, development and evaluation of RSs. However, RSs are not only a technique issue, but also a social issue. More studies are deserved to explore how to well incorporate social impact and social science into the whole life-cycle of RSs.

\textit{New evaluation protocols and metrics.}
As discussed in Section~\ref{what}, building trustworthy RSs triggers new challenges for RS evaluations. On one hand, new technical evaluation protocols and measures are in demand to comprehensively measure the trustworthiness of RSs. On the other hand, new ethical evaluation methods are needed to effectively measure the social impact of RSs.    

\section{Conclusions}
Trustworthy recommender
system (TRS) is an extremely challenging yet demanding topic, which is of both great theoretical and practical values. 
In this paper, we have provided a comprehensive overview of this topic. We comprehensively characterized trustworthy RSs by thoroughly analyzing the various aspects specifying the trustworthiness of RSs. We provided a novel four-stage framework to systematically illustrate trustworthy RSs and analyzed the corresponding challenges in each of the four stages in building trustworthy RSs. We have also described a conceptual framework to support trustworthy RSs and pointed out some future directions in this vibrant area. The research in trustworthy RS is flourishing and it is our hope that this work can provide readers with a comprehensive understanding of the key aspects, main challenges, key techniques when building trustworthy RSs and shade some light on future studies.


\begin{acks}
Prof. Huan Liu from Arizona State University has provided some feedback on the earlier version of this work. This work was supported by Australian Research Council Discovery Project DP200101441.
\end{acks}

\normalem
\small{
\bibliographystyle{ACM-Reference-Format}
\bibliography{Reference.bib}
}

%% file: Main.bbl

\begin{thebibliography}{70}


\ifx \showCODEN    \undefined \def \showCODEN     #1{\unskip}     \fi
\ifx \showDOI      \undefined \def \showDOI       #1{#1}\fi
\ifx \showISBNx    \undefined \def \showISBNx     #1{\unskip}     \fi
\ifx \showISBNxiii \undefined \def \showISBNxiii  #1{\unskip}     \fi
\ifx \showISSN     \undefined \def \showISSN      #1{\unskip}     \fi
\ifx \showLCCN     \undefined \def \showLCCN      #1{\unskip}     \fi
\ifx \shownote     \undefined \def \shownote      #1{#1}          \fi
\ifx \showarticletitle \undefined \def \showarticletitle #1{#1}   \fi
\ifx \showURL      \undefined \def \showURL       {\relax}        \fi
\providecommand\bibfield[2]{#2}
\providecommand\bibinfo[2]{#2}
\providecommand\natexlab[1]{#1}
\providecommand\showeprint[2][]{arXiv:#2}

\bibitem[\protect\citeauthoryear{Abdollahpouri, Adomavicius, Burke, and
  et~al.}{Abdollahpouri et~al\mbox{.}}{2020}]%
        {AbdollahpouriAB20}
\bibfield{author}{\bibinfo{person}{Himan Abdollahpouri},
  \bibinfo{person}{Gediminas Adomavicius}, \bibinfo{person}{Robin Burke}, {and}
  \bibinfo{person}{et al.}} \bibinfo{year}{2020}\natexlab{}.
\newblock \showarticletitle{Multistakeholder recommendation: Survey and
  research directions}.
\newblock \bibinfo{journal}{\emph{User Model. User Adapt. Interact.}}
  \bibinfo{volume}{30}, \bibinfo{number}{1} (\bibinfo{year}{2020}),
  \bibinfo{pages}{127--158}.
\newblock


\bibitem[\protect\citeauthoryear{Adomavicius and Kwon}{Adomavicius and
  Kwon}{2011}]%
        {adomavicius2011improving}
\bibfield{author}{\bibinfo{person}{Gediminas Adomavicius} {and}
  \bibinfo{person}{YoungOk Kwon}.} \bibinfo{year}{2011}\natexlab{}.
\newblock \showarticletitle{Improving aggregate recommendation diversity using
  ranking-based techniques}.
\newblock \bibinfo{journal}{\emph{IEEE Trans Knowl Data Eng}}
  \bibinfo{volume}{24}, \bibinfo{number}{5} (\bibinfo{year}{2011}),
  \bibinfo{pages}{896--911}.
\newblock


\bibitem[\protect\citeauthoryear{Ahmadian, Ahmadi, and Ahmadian}{Ahmadian
  et~al\mbox{.}}{2022}]%
        {ahmadian2022reliable}
\bibfield{author}{\bibinfo{person}{Milad Ahmadian}, \bibinfo{person}{Mahmood
  Ahmadi}, {and} \bibinfo{person}{Sajad Ahmadian}.}
  \bibinfo{year}{2022}\natexlab{}.
\newblock \showarticletitle{A reliable deep representation learning to improve
  trust-aware recommendation systems}.
\newblock \bibinfo{journal}{\emph{Expert Systems with Applications}}
  \bibinfo{volume}{197} (\bibinfo{year}{2022}), \bibinfo{pages}{116697}.
\newblock


\bibitem[\protect\citeauthoryear{Balog, Radlinski, and Arakelyan}{Balog
  et~al\mbox{.}}{2019}]%
        {balog2019transparent}
\bibfield{author}{\bibinfo{person}{Krisztian Balog}, \bibinfo{person}{Filip
  Radlinski}, {and} \bibinfo{person}{Shushan Arakelyan}.}
  \bibinfo{year}{2019}\natexlab{}.
\newblock \showarticletitle{Transparent, scrutable and explainable user models
  for personalized recommendation}. In \bibinfo{booktitle}{\emph{SIGIR}}.
  \bibinfo{pages}{265--274}.
\newblock


\bibitem[\protect\citeauthoryear{Castells, Hurley, and Vargas}{Castells
  et~al\mbox{.}}{2015}]%
        {CastellsHV15}
\bibfield{author}{\bibinfo{person}{Pablo Castells}, \bibinfo{person}{Neil~J.
  Hurley}, {and} \bibinfo{person}{Saul Vargas}.}
  \bibinfo{year}{2015}\natexlab{}.
\newblock \showarticletitle{Novelty and diversity in recommender systems}.
\newblock In \bibinfo{booktitle}{\emph{Recommender Systems Handbook}},
  \bibfield{editor}{\bibinfo{person}{Francesco Ricci}, \bibinfo{person}{Lior
  Rokach}, {and} \bibinfo{person}{Bracha Shapira}} (Eds.).
  \bibinfo{publisher}{Springer}, \bibinfo{pages}{881--918}.
\newblock


\bibitem[\protect\citeauthoryear{Chen, Dong, and et~al.}{Chen
  et~al\mbox{.}}{2021}]%
        {Chen2021Auto}
\bibfield{author}{\bibinfo{person}{Jiawei Chen}, \bibinfo{person}{Hande Dong},
  {and} \bibinfo{person}{et al.}} \bibinfo{year}{2021}\natexlab{}.
\newblock \showarticletitle{AutoDebias: learning to debias for recommendation}.
  In \bibinfo{booktitle}{\emph{SIGIR}}. \bibinfo{pages}{21--30}.
\newblock


\bibitem[\protect\citeauthoryear{Chen, Dong, Wang, Feng, Wang, and He}{Chen
  et~al\mbox{.}}{2023}]%
        {chen2023bias}
\bibfield{author}{\bibinfo{person}{Jiawei Chen}, \bibinfo{person}{Hande Dong},
  \bibinfo{person}{Xiang Wang}, \bibinfo{person}{Fuli Feng},
  \bibinfo{person}{Meng Wang}, {and} \bibinfo{person}{Xiangnan He}.}
  \bibinfo{year}{2023}\natexlab{}.
\newblock \showarticletitle{Bias and debias in recommender system: A survey and
  future directions}.
\newblock \bibinfo{journal}{\emph{ACM Transactions on Information Systems}}
  \bibinfo{volume}{41}, \bibinfo{number}{3} (\bibinfo{year}{2023}),
  \bibinfo{pages}{1--39}.
\newblock


\bibitem[\protect\citeauthoryear{De~Arroyabe, Arranz, Arroyabe, and
  de~Arroyabe}{De~Arroyabe et~al\mbox{.}}{2023}]%
        {de2023cybersecurity}
\bibfield{author}{\bibinfo{person}{Ignacio~Fernandez De~Arroyabe},
  \bibinfo{person}{Carlos~FA Arranz}, \bibinfo{person}{Marta~F Arroyabe}, {and}
  \bibinfo{person}{Juan Carlos~Fernandez de Arroyabe}.}
  \bibinfo{year}{2023}\natexlab{}.
\newblock \showarticletitle{Cybersecurity capabilities and cyber-attacks as
  drivers of investment in cybersecurity systems: A UK survey for 2018 and
  2019}.
\newblock \bibinfo{journal}{\emph{Computers \& Security}}
  \bibinfo{volume}{124} (\bibinfo{year}{2023}), \bibinfo{pages}{102954}.
\newblock


\bibitem[\protect\citeauthoryear{Deldjoo, Noia, and Merra}{Deldjoo
  et~al\mbox{.}}{2021}]%
        {DeldjooNM21}
\bibfield{author}{\bibinfo{person}{Yashar Deldjoo}, \bibinfo{person}{Tommaso~Di
  Noia}, {and} \bibinfo{person}{Felice~Antonio Merra}.}
  \bibinfo{year}{2021}\natexlab{}.
\newblock \showarticletitle{A Survey on adversarial recommender systems: from
  attack/defense strategies to generative adversarial networks}.
\newblock \bibinfo{journal}{\emph{{ACM} Comput. Surv.}} \bibinfo{volume}{54},
  \bibinfo{number}{2} (\bibinfo{year}{2021}), \bibinfo{pages}{1--38}.
\newblock


\bibitem[\protect\citeauthoryear{Di~Noia, Tintarev, Fatourou, and
  Schedl}{Di~Noia et~al\mbox{.}}{2022}]%
        {di2022recommender}
\bibfield{author}{\bibinfo{person}{Tommaso Di~Noia}, \bibinfo{person}{Nava
  Tintarev}, \bibinfo{person}{Panagiota Fatourou}, {and}
  \bibinfo{person}{Markus Schedl}.} \bibinfo{year}{2022}\natexlab{}.
\newblock \showarticletitle{Recommender systems under European AI regulations}.
\newblock \bibinfo{journal}{\emph{Commun. ACM}} \bibinfo{volume}{65},
  \bibinfo{number}{4} (\bibinfo{year}{2022}), \bibinfo{pages}{69--73}.
\newblock


\bibitem[\protect\citeauthoryear{Dictionaries}{Dictionaries}{2022}]%
        {trustworthy}
\bibfield{author}{\bibinfo{person}{O.~L. Dictionaries}.}
  \bibinfo{year}{2022}\natexlab{}.
\newblock \bibinfo{title}{Definition of trustworthy}.
\newblock
\newblock
\newblock
\shownote{\url{https://www.oxfordlearnersdictionaries.com/definition/american
  english/trustworthy}.}


\bibitem[\protect\citeauthoryear{Donahue and Simonyan}{Donahue and
  Simonyan}{2019}]%
        {donahue2019large}
\bibfield{author}{\bibinfo{person}{Jeff Donahue} {and} \bibinfo{person}{Karen
  Simonyan}.} \bibinfo{year}{2019}\natexlab{}.
\newblock \showarticletitle{Large scale adversarial representation learning}.
  In \bibinfo{booktitle}{\emph{NIPS}}. \bibinfo{pages}{10542–10552}.
\newblock


\bibitem[\protect\citeauthoryear{Dong, Yuan, Yao, Wang, Xu, and Zhu}{Dong
  et~al\mbox{.}}{2022}]%
        {dong2022survey}
\bibfield{author}{\bibinfo{person}{Manqing Dong}, \bibinfo{person}{Feng Yuan},
  \bibinfo{person}{Lina Yao}, \bibinfo{person}{Xianzhi Wang},
  \bibinfo{person}{Xiwei Xu}, {and} \bibinfo{person}{Liming Zhu}.}
  \bibinfo{year}{2022}\natexlab{}.
\newblock \showarticletitle{A survey for trust-aware recommender systems: a
  deep learning perspective}.
\newblock \bibinfo{journal}{\emph{Knowledge-Based Systems}}
  (\bibinfo{year}{2022}), \bibinfo{pages}{108954}.
\newblock


\bibitem[\protect\citeauthoryear{Elahi, Jannach, Skj{\ae}rven, and
  et~al.}{Elahi et~al\mbox{.}}{2021}]%
        {elahi2021towards}
\bibfield{author}{\bibinfo{person}{Mehdi Elahi}, \bibinfo{person}{Dietmar
  Jannach}, \bibinfo{person}{Lars Skj{\ae}rven}, {and} \bibinfo{person}{et
  al.}} \bibinfo{year}{2021}\natexlab{}.
\newblock \showarticletitle{Towards responsible media recommendation}.
\newblock \bibinfo{journal}{\emph{AI and Ethics}} (\bibinfo{year}{2021}),
  \bibinfo{pages}{1--12}.
\newblock


\bibitem[\protect\citeauthoryear{Ge, Liu, Fu, and et~al.}{Ge
  et~al\mbox{.}}{2022a}]%
        {ge2022survey}
\bibfield{author}{\bibinfo{person}{Yingqiang Ge}, \bibinfo{person}{Shuchang
  Liu}, \bibinfo{person}{Zuohui Fu}, {and} \bibinfo{person}{et al.}}
  \bibinfo{year}{2022}\natexlab{a}.
\newblock \showarticletitle{A survey on trustworthy recommender systems}.
\newblock \bibinfo{journal}{\emph{arXiv preprint arXiv:2207.12515}}
  (\bibinfo{year}{2022}), \bibinfo{pages}{1--43}.
\newblock


\bibitem[\protect\citeauthoryear{Ge, Tan, Zhu, Xia, Luo, and et~al.}{Ge
  et~al\mbox{.}}{2022b}]%
        {ge2022explainable}
\bibfield{author}{\bibinfo{person}{Yingqiang Ge}, \bibinfo{person}{Juntao Tan},
  \bibinfo{person}{Yan Zhu}, \bibinfo{person}{Yinglong Xia},
  \bibinfo{person}{Jiebo Luo}, {and} \bibinfo{person}{et al.}}
  \bibinfo{year}{2022}\natexlab{b}.
\newblock \showarticletitle{Explainable fairness in recommendation}. In
  \bibinfo{booktitle}{\emph{Proceedings of the 45th International ACM SIGIR
  Conference on Research and Development in Information Retrieval}}.
  \bibinfo{pages}{681--691}.
\newblock


\bibitem[\protect\citeauthoryear{Giansiracusa}{Giansiracusa}{2021}]%
        {giansiracusa2021tools}
\bibfield{author}{\bibinfo{person}{Noah Giansiracusa}.}
  \bibinfo{year}{2021}\natexlab{}.
\newblock \showarticletitle{Tools for truth}.
\newblock In \bibinfo{booktitle}{\emph{How Algorithms Create and Prevent Fake
  News}}. \bibinfo{publisher}{Springer}, \bibinfo{pages}{217--229}.
\newblock


\bibitem[\protect\citeauthoryear{Goldberg, Nichols, Oki, and Terry}{Goldberg
  et~al\mbox{.}}{1992}]%
        {goldberg1992using}
\bibfield{author}{\bibinfo{person}{David Goldberg}, \bibinfo{person}{David
  Nichols}, \bibinfo{person}{Brian~M Oki}, {and} \bibinfo{person}{Douglas
  Terry}.} \bibinfo{year}{1992}\natexlab{}.
\newblock \showarticletitle{Using collaborative filtering to weave an
  information tapestry}.
\newblock \bibinfo{journal}{\emph{Commun. ACM}} \bibinfo{volume}{35},
  \bibinfo{number}{12} (\bibinfo{year}{1992}), \bibinfo{pages}{61--70}.
\newblock


\bibitem[\protect\citeauthoryear{Gunawardana and Shani}{Gunawardana and
  Shani}{2015}]%
        {GunawardanaS15}
\bibfield{author}{\bibinfo{person}{Asela Gunawardana} {and}
  \bibinfo{person}{Guy Shani}.} \bibinfo{year}{2015}\natexlab{}.
\newblock \showarticletitle{Evaluating Recommender Systems}.
\newblock In \bibinfo{booktitle}{\emph{Recommender Systems Handbook}},
  \bibfield{editor}{\bibinfo{person}{Francesco Ricci}, \bibinfo{person}{Lior
  Rokach}, {and} \bibinfo{person}{Bracha Shapira}} (Eds.).
  \bibinfo{publisher}{Springer}, \bibinfo{pages}{265--308}.
\newblock


\bibitem[\protect\citeauthoryear{Gunes, Kaleli, Bilge, and Polat}{Gunes
  et~al\mbox{.}}{2014}]%
        {gunes2014shilling}
\bibfield{author}{\bibinfo{person}{Ihsan Gunes}, \bibinfo{person}{Cihan
  Kaleli}, \bibinfo{person}{Alper Bilge}, {and} \bibinfo{person}{Huseyin
  Polat}.} \bibinfo{year}{2014}\natexlab{}.
\newblock \showarticletitle{Shilling attacks against recommender systems: a
  comprehensive survey}.
\newblock \bibinfo{journal}{\emph{Artificial Intelligence Review}}
  \bibinfo{volume}{42}, \bibinfo{number}{4} (\bibinfo{year}{2014}),
  \bibinfo{pages}{767--799}.
\newblock


\bibitem[\protect\citeauthoryear{Hazrati and Ricci}{Hazrati and Ricci}{2022}]%
        {HazratiR22}
\bibfield{author}{\bibinfo{person}{Naieme Hazrati} {and}
  \bibinfo{person}{Francesco Ricci}.} \bibinfo{year}{2022}\natexlab{}.
\newblock \showarticletitle{Recommender systems effect on the evolution of
  users' choices distribution}.
\newblock \bibinfo{journal}{\emph{Inf. Process. Manag.}} \bibinfo{volume}{59},
  \bibinfo{number}{1} (\bibinfo{year}{2022}), \bibinfo{pages}{102766}.
\newblock


\bibitem[\protect\citeauthoryear{He, Xu, Huang, and Deng}{He
  et~al\mbox{.}}{2005}]%
        {he2005fp}
\bibfield{author}{\bibinfo{person}{Zengyou He}, \bibinfo{person}{Xiaofei Xu},
  \bibinfo{person}{Zhexue~Joshua Huang}, {and} \bibinfo{person}{Shengchun
  Deng}.} \bibinfo{year}{2005}\natexlab{}.
\newblock \showarticletitle{FP-outlier: frequent pattern based outlier
  detection}.
\newblock \bibinfo{journal}{\emph{Computer Science and Information Systems}}
  \bibinfo{volume}{2}, \bibinfo{number}{1} (\bibinfo{year}{2005}),
  \bibinfo{pages}{103--118}.
\newblock


\bibitem[\protect\citeauthoryear{Huang, Siegel, and Madnick}{Huang
  et~al\mbox{.}}{2018}]%
        {huang2018systematically}
\bibfield{author}{\bibinfo{person}{Keman Huang}, \bibinfo{person}{Michael
  Siegel}, {and} \bibinfo{person}{Stuart Madnick}.}
  \bibinfo{year}{2018}\natexlab{}.
\newblock \showarticletitle{Systematically understanding the cyber attack
  business: a survey}.
\newblock \bibinfo{journal}{\emph{Comput. Surveys}} \bibinfo{volume}{51},
  \bibinfo{number}{4} (\bibinfo{year}{2018}), \bibinfo{pages}{1--36}.
\newblock


\bibitem[\protect\citeauthoryear{Imran, Yin, Chen, Nguyen, Zhou, and
  Zheng}{Imran et~al\mbox{.}}{2023}]%
        {imran2023refrs}
\bibfield{author}{\bibinfo{person}{Mubashir Imran}, \bibinfo{person}{Hongzhi
  Yin}, \bibinfo{person}{Tong Chen}, \bibinfo{person}{Quoc Viet~Hung Nguyen},
  \bibinfo{person}{Alexander Zhou}, {and} \bibinfo{person}{Kai Zheng}.}
  \bibinfo{year}{2023}\natexlab{}.
\newblock \showarticletitle{ReFRS: Resource-efficient federated recommender
  system for dynamic and diversified user preferences}.
\newblock \bibinfo{journal}{\emph{ACM Transactions on Information Systems}}
  \bibinfo{volume}{41}, \bibinfo{number}{3} (\bibinfo{year}{2023}),
  \bibinfo{pages}{1--30}.
\newblock


\bibitem[\protect\citeauthoryear{Jannach, Pu, Ricci, and Zanker}{Jannach
  et~al\mbox{.}}{2021}]%
        {JannachPRZ21}
\bibfield{author}{\bibinfo{person}{Dietmar Jannach}, \bibinfo{person}{Pearl
  Pu}, \bibinfo{person}{Francesco Ricci}, {and} \bibinfo{person}{Markus
  Zanker}.} \bibinfo{year}{2021}\natexlab{}.
\newblock \showarticletitle{Recommender systems: past, present, future}.
\newblock \bibinfo{journal}{\emph{{AI} Mag.}} \bibinfo{volume}{42},
  \bibinfo{number}{3} (\bibinfo{year}{2021}), \bibinfo{pages}{3--6}.
\newblock


\bibitem[\protect\citeauthoryear{Jannach, Resnick, Tuzhilin, and
  Zanker}{Jannach et~al\mbox{.}}{2016}]%
        {jannach2016recommender}
\bibfield{author}{\bibinfo{person}{Dietmar Jannach}, \bibinfo{person}{Paul
  Resnick}, \bibinfo{person}{Alexander Tuzhilin}, {and} \bibinfo{person}{Markus
  Zanker}.} \bibinfo{year}{2016}\natexlab{}.
\newblock \showarticletitle{Recommender systems—beyond matrix completion}.
\newblock \bibinfo{journal}{\emph{Commun. ACM}} \bibinfo{volume}{59},
  \bibinfo{number}{11} (\bibinfo{year}{2016}), \bibinfo{pages}{94--102}.
\newblock


\bibitem[\protect\citeauthoryear{Jha, Gaur, Ranjan, and Thakur}{Jha
  et~al\mbox{.}}{2021}]%
        {jha2021survey}
\bibfield{author}{\bibinfo{person}{Govind~Kumar Jha}, \bibinfo{person}{Manish
  Gaur}, \bibinfo{person}{Preetish Ranjan}, {and} \bibinfo{person}{Hardeo~Kumar
  Thakur}.} \bibinfo{year}{2021}\natexlab{}.
\newblock \showarticletitle{A survey on trustworthy model of recommender
  system}.
\newblock \bibinfo{journal}{\emph{International Journal of System Assurance
  Engineering and Management}} (\bibinfo{year}{2021}), \bibinfo{pages}{1--18}.
\newblock


\bibitem[\protect\citeauthoryear{Jin, Wang, Zhang, Zheng, Ding, Xia, and
  Pan}{Jin et~al\mbox{.}}{2023}]%
        {jin2023survey}
\bibfield{author}{\bibinfo{person}{Di Jin}, \bibinfo{person}{Luzhi Wang},
  \bibinfo{person}{He Zhang}, \bibinfo{person}{Yizhen Zheng},
  \bibinfo{person}{Weiping Ding}, \bibinfo{person}{Feng Xia}, {and}
  \bibinfo{person}{Shirui Pan}.} \bibinfo{year}{2023}\natexlab{}.
\newblock \showarticletitle{A Survey on Fairness-aware Recommender Systems}.
\newblock \bibinfo{journal}{\emph{arXiv preprint arXiv:2306.00403}}
  (\bibinfo{year}{2023}).
\newblock


\bibitem[\protect\citeauthoryear{Konstan and Terveen}{Konstan and
  Terveen}{2021}]%
        {KonstanT21}
\bibfield{author}{\bibinfo{person}{Joseph~A. Konstan} {and}
  \bibinfo{person}{Loren~G. Terveen}.} \bibinfo{year}{2021}\natexlab{}.
\newblock \showarticletitle{Human-Centered Recommender Systems: Origins,
  Advances, Challenges, and Opportunities}.
\newblock \bibinfo{journal}{\emph{{AI} Mag.}} \bibinfo{volume}{42},
  \bibinfo{number}{3} (\bibinfo{year}{2021}), \bibinfo{pages}{31--42}.
\newblock


\bibitem[\protect\citeauthoryear{Lam, Frankowski, and et~al.}{Lam
  et~al\mbox{.}}{2006}]%
        {lam2006you}
\bibfield{author}{\bibinfo{person}{Shyong~K Lam}, \bibinfo{person}{Dan
  Frankowski}, {and} \bibinfo{person}{et al.}} \bibinfo{year}{2006}\natexlab{}.
\newblock \showarticletitle{Do you trust your recommendations? An exploration
  of security and privacy issues in recommender systems}. In
  \bibinfo{booktitle}{\emph{ETRICS}}. \bibinfo{pages}{14--29}.
\newblock


\bibitem[\protect\citeauthoryear{Leonhardt and et~al.}{Leonhardt and
  et~al.}{2018}]%
        {leonhardt2018user}
\bibfield{author}{\bibinfo{person}{Jurek Leonhardt} {and} \bibinfo{person}{et
  al.}} \bibinfo{year}{2018}\natexlab{}.
\newblock \showarticletitle{User fairness in recommender systems}. In
  \bibinfo{booktitle}{\emph{WWW Companion}}. \bibinfo{pages}{101--102}.
\newblock


\bibitem[\protect\citeauthoryear{Li, Zhang, and Chen}{Li et~al\mbox{.}}{2023}]%
        {li2023personalized}
\bibfield{author}{\bibinfo{person}{Lei Li}, \bibinfo{person}{Yongfeng Zhang},
  {and} \bibinfo{person}{Li Chen}.} \bibinfo{year}{2023}\natexlab{}.
\newblock \showarticletitle{Personalized prompt learning for explainable
  recommendation}.
\newblock \bibinfo{journal}{\emph{ACM Transactions on Information Systems}}
  \bibinfo{volume}{41}, \bibinfo{number}{4} (\bibinfo{year}{2023}),
  \bibinfo{pages}{1--26}.
\newblock


\bibitem[\protect\citeauthoryear{Liu, Yang, Fan, Peng, and Yu}{Liu
  et~al\mbox{.}}{2022}]%
        {liu2022federated}
\bibfield{author}{\bibinfo{person}{Zhiwei Liu}, \bibinfo{person}{Liangwei
  Yang}, \bibinfo{person}{Ziwei Fan}, \bibinfo{person}{Hao Peng}, {and}
  \bibinfo{person}{Philip~S Yu}.} \bibinfo{year}{2022}\natexlab{}.
\newblock \showarticletitle{Federated social recommendation with graph neural
  network}.
\newblock \bibinfo{journal}{\emph{ACM Transactions on Intelligent Systems and
  Technology (TIST)}} \bibinfo{volume}{13}, \bibinfo{number}{4}
  (\bibinfo{year}{2022}), \bibinfo{pages}{1--24}.
\newblock


\bibitem[\protect\citeauthoryear{Lu, Meng, Wang, Zhang, Zhang, Ouyang, and
  Zhang}{Lu et~al\mbox{.}}{2019}]%
        {lu2019graph}
\bibfield{author}{\bibinfo{person}{Wenpeng Lu}, \bibinfo{person}{Fanqing Meng},
  \bibinfo{person}{Shoujin Wang}, \bibinfo{person}{Guoqiang Zhang},
  \bibinfo{person}{Xu Zhang}, \bibinfo{person}{Antai Ouyang}, {and}
  \bibinfo{person}{Xiaodong Zhang}.} \bibinfo{year}{2019}\natexlab{}.
\newblock \showarticletitle{Graph-Based Chinese Word Sense Disambiguation with
  Multi-Knowledge Integration.}
\newblock \bibinfo{journal}{\emph{Computers, Materials \& Continua}}
  \bibinfo{volume}{61}, \bibinfo{number}{1} (\bibinfo{year}{2019}).
\newblock


\bibitem[\protect\citeauthoryear{Lu, Tsao, Matsuda, and Hori}{Lu
  et~al\mbox{.}}{2013}]%
        {lu2013speech}
\bibfield{author}{\bibinfo{person}{Xugang Lu}, \bibinfo{person}{Yu Tsao},
  \bibinfo{person}{Shigeki Matsuda}, {and} \bibinfo{person}{Chiori Hori}.}
  \bibinfo{year}{2013}\natexlab{}.
\newblock \showarticletitle{Speech enhancement based on deep denoising
  autoencoder}. In \bibinfo{booktitle}{\emph{Interspeech}},
  Vol.~\bibinfo{volume}{2013}. \bibinfo{pages}{436--440}.
\newblock


\bibitem[\protect\citeauthoryear{Lyu, Yin, Liu, Liu, Liu, and Deng}{Lyu
  et~al\mbox{.}}{2021}]%
        {lyu2021reliable}
\bibfield{author}{\bibinfo{person}{Yanzhang Lyu}, \bibinfo{person}{Hongzhi
  Yin}, \bibinfo{person}{Jun Liu}, \bibinfo{person}{Mengyue Liu},
  \bibinfo{person}{Huan Liu}, {and} \bibinfo{person}{Shizhuo Deng}.}
  \bibinfo{year}{2021}\natexlab{}.
\newblock \showarticletitle{Reliable recommendation with review-level
  explanations}. In \bibinfo{booktitle}{\emph{ICDE}}.
  \bibinfo{pages}{1548--1558}.
\newblock


\bibitem[\protect\citeauthoryear{Ma, Zhou, Cui, Yang, and Zhu}{Ma
  et~al\mbox{.}}{2019}]%
        {ma2019learning}
\bibfield{author}{\bibinfo{person}{Jianxin Ma}, \bibinfo{person}{Chang Zhou},
  \bibinfo{person}{Peng Cui}, \bibinfo{person}{Hongxia Yang}, {and}
  \bibinfo{person}{Wenwu Zhu}.} \bibinfo{year}{2019}\natexlab{}.
\newblock \showarticletitle{Learning disentangled representations for
  recommendation}. In \bibinfo{booktitle}{\emph{NIPS}}.
\newblock


\bibitem[\protect\citeauthoryear{Massa and Avesani}{Massa and Avesani}{2007}]%
        {massa2007trust}
\bibfield{author}{\bibinfo{person}{Paolo Massa} {and} \bibinfo{person}{Paolo
  Avesani}.} \bibinfo{year}{2007}\natexlab{}.
\newblock \showarticletitle{Trust-aware recommender systems}. In
  \bibinfo{booktitle}{\emph{RecSys}}. \bibinfo{pages}{17--24}.
\newblock


\bibitem[\protect\citeauthoryear{Mehrabi, Morstatter, Saxena, and
  et~al.}{Mehrabi et~al\mbox{.}}{2021}]%
        {mehrabi2021survey}
\bibfield{author}{\bibinfo{person}{Ninareh Mehrabi}, \bibinfo{person}{Fred
  Morstatter}, \bibinfo{person}{Nripsuta Saxena}, {and} \bibinfo{person}{et
  al.}} \bibinfo{year}{2021}\natexlab{}.
\newblock \showarticletitle{A survey on bias and fairness in machine learning}.
\newblock \bibinfo{journal}{\emph{Comput. Surveys}} \bibinfo{volume}{54},
  \bibinfo{number}{6} (\bibinfo{year}{2021}), \bibinfo{pages}{1--35}.
\newblock


\bibitem[\protect\citeauthoryear{Mobasher, Burke, Bhaumik, and et~al.}{Mobasher
  et~al\mbox{.}}{2007}]%
        {mobasher2007toward}
\bibfield{author}{\bibinfo{person}{Bamshad Mobasher}, \bibinfo{person}{Robin
  Burke}, \bibinfo{person}{Runa Bhaumik}, {and} \bibinfo{person}{et al.}}
  \bibinfo{year}{2007}\natexlab{}.
\newblock \showarticletitle{Toward trustworthy recommender systems: an analysis
  of attack models and algorithm robustness}.
\newblock \bibinfo{journal}{\emph{ACM Transactions on Internet Technology}}
  \bibinfo{volume}{7}, \bibinfo{number}{4} (\bibinfo{year}{2007}),
  \bibinfo{pages}{23--es}.
\newblock


\bibitem[\protect\citeauthoryear{Ning, Desrosiers, and et~al.}{Ning
  et~al\mbox{.}}{2015}]%
        {NingDK15}
\bibfield{author}{\bibinfo{person}{Xia Ning}, \bibinfo{person}{Christian
  Desrosiers}, {and} \bibinfo{person}{et al.}} \bibinfo{year}{2015}\natexlab{}.
\newblock \showarticletitle{A comprehensive survey of neighborhood-based
  recommendation methods}.
\newblock In \bibinfo{booktitle}{\emph{Recommender Systems Handbook}},
  \bibfield{editor}{\bibinfo{person}{Francesco Ricci}, \bibinfo{person}{Lior
  Rokach}, {and} \bibinfo{person}{Bracha Shapira}} (Eds.).
  \bibinfo{pages}{37--76}.
\newblock


\bibitem[\protect\citeauthoryear{of~China}{of~China}{2022}]%
        {TRS_law}
\bibfield{author}{\bibinfo{person}{Cyberspace~Administration of China}.}
  \bibinfo{year}{2022}\natexlab{}.
\newblock \bibinfo{title}{Regulation rules on the recommendation algorithm for
  internet information service}.
\newblock
\newblock
\newblock
\shownote{\url{http://www.cac.gov.cn/2022-01/04/c_1642894606364259.htm}.}


\bibitem[\protect\citeauthoryear{Pang, Shen, Cao, and et~al.}{Pang
  et~al\mbox{.}}{2021}]%
        {pang2021deep}
\bibfield{author}{\bibinfo{person}{Guansong Pang}, \bibinfo{person}{Chunhua
  Shen}, \bibinfo{person}{Longbing Cao}, {and} \bibinfo{person}{et al.}}
  \bibinfo{year}{2021}\natexlab{}.
\newblock \showarticletitle{Deep learning for anomaly detection: a review}.
\newblock \bibinfo{journal}{\emph{Comput. Surveys}} \bibinfo{volume}{54},
  \bibinfo{number}{2} (\bibinfo{year}{2021}), \bibinfo{pages}{1--38}.
\newblock


\bibitem[\protect\citeauthoryear{Ricci, Rokach, and Shapira}{Ricci
  et~al\mbox{.}}{2022}]%
        {2022rsh}
\bibfield{editor}{\bibinfo{person}{Francesco Ricci}, \bibinfo{person}{Lior
  Rokach}, {and} \bibinfo{person}{Bracha Shapira}} (Eds.).
  \bibinfo{year}{2022}\natexlab{}.
\newblock \bibinfo{booktitle}{\emph{Recommender Systems Handbook}}.
\newblock \bibinfo{publisher}{Springer}.
\newblock


\bibitem[\protect\citeauthoryear{Rong}{Rong}{2014}]%
        {rong2014word2vec}
\bibfield{author}{\bibinfo{person}{Xin Rong}.} \bibinfo{year}{2014}\natexlab{}.
\newblock \showarticletitle{Word2vec parameter learning explained}.
\newblock \bibinfo{journal}{\emph{arXiv preprint arXiv:1411.2738}}
  (\bibinfo{year}{2014}).
\newblock


\bibitem[\protect\citeauthoryear{Schnabel, Swaminathan, and et~al.}{Schnabel
  et~al\mbox{.}}{2016}]%
        {SchnabelSSCJ16}
\bibfield{author}{\bibinfo{person}{Tobias Schnabel}, \bibinfo{person}{Adith
  Swaminathan}, {and} \bibinfo{person}{et al.}}
  \bibinfo{year}{2016}\natexlab{}.
\newblock \showarticletitle{Recommendations as treatments: debiasing learning
  and evaluation}. In \bibinfo{booktitle}{\emph{ICML}}.
  \bibinfo{pages}{1670--1679}.
\newblock


\bibitem[\protect\citeauthoryear{Shi and et~al.}{Shi and et~al.}{2016}]%
        {shi2016edge}
\bibfield{author}{\bibinfo{person}{Weisong Shi} {and} \bibinfo{person}{et al.}}
  \bibinfo{year}{2016}\natexlab{}.
\newblock \showarticletitle{Edge computing: vision and challenges}.
\newblock \bibinfo{journal}{\emph{IEEE Internet of Things Journal}}
  \bibinfo{volume}{3}, \bibinfo{number}{5} (\bibinfo{year}{2016}),
  \bibinfo{pages}{637--646}.
\newblock


\bibitem[\protect\citeauthoryear{Shin, Kim, and et~al.}{Shin
  et~al\mbox{.}}{2018}]%
        {shin2018privacy}
\bibfield{author}{\bibinfo{person}{Hyejin Shin}, \bibinfo{person}{Sungwook
  Kim}, {and} \bibinfo{person}{et al.}} \bibinfo{year}{2018}\natexlab{}.
\newblock \showarticletitle{Privacy enhanced matrix factorization for
  recommendation with local differential privacy}.
\newblock \bibinfo{journal}{\emph{IEEE Trans Knowl Data Eng}}
  \bibinfo{volume}{30}, \bibinfo{number}{9} (\bibinfo{year}{2018}),
  \bibinfo{pages}{1770--1782}.
\newblock


\bibitem[\protect\citeauthoryear{Tian, Sun, Poole, Krishnan, Schmid, and
  Isola}{Tian et~al\mbox{.}}{2020}]%
        {tian2020makes}
\bibfield{author}{\bibinfo{person}{Yonglong Tian}, \bibinfo{person}{Chen Sun},
  \bibinfo{person}{Ben Poole}, \bibinfo{person}{Dilip Krishnan},
  \bibinfo{person}{Cordelia Schmid}, {and} \bibinfo{person}{Phillip Isola}.}
  \bibinfo{year}{2020}\natexlab{}.
\newblock \showarticletitle{What makes for good views for contrastive
  learning?}. In \bibinfo{booktitle}{\emph{NIPS}}. \bibinfo{pages}{6827--6839}.
\newblock


\bibitem[\protect\citeauthoryear{Van~Maaten and et~al.}{Van~Maaten and
  et~al.}{2008}]%
        {van2008visualizing}
\bibfield{author}{\bibinfo{person}{Laurens Van~Maaten} {and}
  \bibinfo{person}{et al.}} \bibinfo{year}{2008}\natexlab{}.
\newblock \showarticletitle{Visualizing data using t-SNE.}
\newblock \bibinfo{journal}{\emph{Journal of Machine Learning Research}}
  \bibinfo{volume}{9}, \bibinfo{number}{11} (\bibinfo{year}{2008}).
\newblock


\bibitem[\protect\citeauthoryear{Wang, Cao, Wang, Sheng, Orgun, and Lian}{Wang
  et~al\mbox{.}}{2021}]%
        {wang2021survey}
\bibfield{author}{\bibinfo{person}{Shoujin Wang}, \bibinfo{person}{Longbing
  Cao}, \bibinfo{person}{Yan Wang}, \bibinfo{person}{Quan~Z Sheng},
  \bibinfo{person}{Mehmet~A Orgun}, {and} \bibinfo{person}{Defu Lian}.}
  \bibinfo{year}{2021}\natexlab{}.
\newblock \showarticletitle{A survey on session-based recommender systems}.
\newblock \bibinfo{journal}{\emph{Comput. Surveys}} \bibinfo{volume}{54},
  \bibinfo{number}{7} (\bibinfo{year}{2021}), \bibinfo{pages}{1--38}.
\newblock


\bibitem[\protect\citeauthoryear{Wang, Wang, Sivrikaya, Albayrak, and
  Anelli}{Wang et~al\mbox{.}}{2023b}]%
        {wang2023data}
\bibfield{author}{\bibinfo{person}{Shoujin Wang}, \bibinfo{person}{Yan Wang},
  \bibinfo{person}{Fikret Sivrikaya}, \bibinfo{person}{Sahin Albayrak}, {and}
  \bibinfo{person}{Vito~Walter Anelli}.} \bibinfo{year}{2023}\natexlab{b}.
\newblock \showarticletitle{Data science for next-generation recommender
  systems}.
\newblock \bibinfo{journal}{\emph{International Journal of Data Science and
  Analytics}} (\bibinfo{year}{2023}).
\newblock
\urldef\tempurl%
\url{https://doi.org/10.1007/s41060-023-00404-w}
\showDOI{\tempurl}


\bibitem[\protect\citeauthoryear{Wang, Xu, Zhang, Wang, and et~al.}{Wang
  et~al\mbox{.}}{2022}]%
        {wang2022veracity}
\bibfield{author}{\bibinfo{person}{Shoujin Wang}, \bibinfo{person}{Xiaofei Xu},
  \bibinfo{person}{Xiuzhen Zhang}, \bibinfo{person}{Yan Wang}, {and}
  \bibinfo{person}{et al.}} \bibinfo{year}{2022}\natexlab{}.
\newblock \showarticletitle{Veracity-aware and event-driven personalized news
  recommendation for fake news mitigation}. In \bibinfo{booktitle}{\emph{WWW}}.
  \bibinfo{pages}{3673--3684}.
\newblock


\bibitem[\protect\citeauthoryear{Wang, Li, and Liu}{Wang et~al\mbox{.}}{2015}]%
        {wang2015social}
\bibfield{author}{\bibinfo{person}{Yan Wang}, \bibinfo{person}{Lei Li}, {and}
  \bibinfo{person}{Guanfeng Liu}.} \bibinfo{year}{2015}\natexlab{}.
\newblock \showarticletitle{Social context-aware trust inference for trust
  enhancement in social network based recommendations on service providers}.
\newblock \bibinfo{journal}{\emph{World Wide Web}} \bibinfo{volume}{18},
  \bibinfo{number}{1} (\bibinfo{year}{2015}), \bibinfo{pages}{159--184}.
\newblock


\bibitem[\protect\citeauthoryear{Wang and Lin}{Wang and Lin}{2006}]%
        {wang2006trust}
\bibfield{author}{\bibinfo{person}{Yan Wang} {and} \bibinfo{person}{Fu-ren
  Lin}.} \bibinfo{year}{2006}\natexlab{}.
\newblock \showarticletitle{Trust and risk evaluation of transactions with
  different amounts in peer-to-peer e-commerce environments}. In
  \bibinfo{booktitle}{\emph{ICEBE}}. \bibinfo{pages}{102--109}.
\newblock


\bibitem[\protect\citeauthoryear{Wang, Ma, Zhang, Liu, and Ma}{Wang
  et~al\mbox{.}}{2023a}]%
        {wang2023survey}
\bibfield{author}{\bibinfo{person}{Yifan Wang}, \bibinfo{person}{Weizhi Ma},
  \bibinfo{person}{Min Zhang}, \bibinfo{person}{Yiqun Liu}, {and}
  \bibinfo{person}{Shaoping Ma}.} \bibinfo{year}{2023}\natexlab{a}.
\newblock \showarticletitle{A survey on the fairness of recommender systems}.
\newblock \bibinfo{journal}{\emph{ACM Transactions on Information Systems}}
  \bibinfo{volume}{41}, \bibinfo{number}{3} (\bibinfo{year}{2023}),
  \bibinfo{pages}{1--43}.
\newblock


\bibitem[\protect\citeauthoryear{Wang, Wong, and et~al.}{Wang
  et~al\mbox{.}}{2008}]%
        {wang2008evaluating}
\bibfield{author}{\bibinfo{person}{Yan Wang}, \bibinfo{person}{Duncan~S Wong},
  {and} \bibinfo{person}{et al.}} \bibinfo{year}{2008}\natexlab{}.
\newblock \showarticletitle{Evaluating transaction trust and risk levels in
  peer-to-peer e-commerce environments}.
\newblock \bibinfo{journal}{\emph{Inf Syst E-Bus Manag}} \bibinfo{volume}{6},
  \bibinfo{number}{1} (\bibinfo{year}{2008}), \bibinfo{pages}{25--48}.
\newblock


\bibitem[\protect\citeauthoryear{Wang and et~al.}{Wang and et~al.}{2021}]%
        {wang2021counterfactual}
\bibfield{author}{\bibinfo{person}{Zhenlei Wang} {and} \bibinfo{person}{et
  al.}} \bibinfo{year}{2021}\natexlab{}.
\newblock \showarticletitle{Counterfactual data-augmented sequential
  recommendation}. In \bibinfo{booktitle}{\emph{SIGIR}}.
  \bibinfo{pages}{347--356}.
\newblock


\bibitem[\protect\citeauthoryear{Wu, He, and et~al.}{Wu et~al\mbox{.}}{2022}]%
        {wu2022survey}
\bibfield{author}{\bibinfo{person}{Le Wu}, \bibinfo{person}{Xiangnan He}, {and}
  \bibinfo{person}{et al.}} \bibinfo{year}{2022}\natexlab{}.
\newblock \showarticletitle{A survey on accuracy-oriented neural
  recommendation: from collaborative filtering to information-rich
  recommendation}.
\newblock \bibinfo{journal}{\emph{IEEE Trans Knowl Data Eng}}
  (\bibinfo{year}{2022}).
\newblock


\bibitem[\protect\citeauthoryear{Wu and et~al.}{Wu and et~al.}{2012}]%
        {wu2012hysad}
\bibfield{author}{\bibinfo{person}{Zhiang Wu} {and} \bibinfo{person}{et al.}}
  \bibinfo{year}{2012}\natexlab{}.
\newblock \showarticletitle{HySAD: A semi-supervised hybrid shilling attack
  detector for trustworthy product recommendation}. In
  \bibinfo{booktitle}{\emph{KDD}}. \bibinfo{pages}{985--993}.
\newblock


\bibitem[\protect\citeauthoryear{Xiao, Min, Yongfeng, Zhaoquan, Yiqun, and
  et~al.}{Xiao et~al\mbox{.}}{2017}]%
        {xiao2017fairness}
\bibfield{author}{\bibinfo{person}{Lin Xiao}, \bibinfo{person}{Zhang Min},
  \bibinfo{person}{Zhang Yongfeng}, \bibinfo{person}{Gu Zhaoquan},
  \bibinfo{person}{Liu Yiqun}, {and} \bibinfo{person}{et al.}}
  \bibinfo{year}{2017}\natexlab{}.
\newblock \showarticletitle{Fairness-aware group recommendation with
  pareto-efficiency}. In \bibinfo{booktitle}{\emph{RecSys}}.
  \bibinfo{pages}{107--115}.
\newblock


\bibitem[\protect\citeauthoryear{Xiong, Ren, and et~al.}{Xiong
  et~al\mbox{.}}{2020}]%
        {xiong2020loco}
\bibfield{author}{\bibinfo{person}{Yuwen Xiong}, \bibinfo{person}{Mengye Ren},
  {and} \bibinfo{person}{et al.}} \bibinfo{year}{2020}\natexlab{}.
\newblock \showarticletitle{Loco: local contrastive representation learning}.
  In \bibinfo{booktitle}{\emph{NIPS}}. \bibinfo{pages}{11142--11153}.
\newblock


\bibitem[\protect\citeauthoryear{Yang, Liu, Cheng, Kang, Chen, and Yu}{Yang
  et~al\mbox{.}}{2019}]%
        {yang2019federated}
\bibfield{author}{\bibinfo{person}{Qiang Yang}, \bibinfo{person}{Yang Liu},
  \bibinfo{person}{Yong Cheng}, \bibinfo{person}{Yan Kang},
  \bibinfo{person}{Tianjian Chen}, {and} \bibinfo{person}{Han Yu}.}
  \bibinfo{year}{2019}\natexlab{}.
\newblock \showarticletitle{Federated learning}.
\newblock \bibinfo{journal}{\emph{Synthesis Lectures on Artificial Intelligence
  and Machine Learning}} \bibinfo{volume}{13}, \bibinfo{number}{3}
  (\bibinfo{year}{2019}), \bibinfo{pages}{1--207}.
\newblock


\bibitem[\protect\citeauthoryear{Zhang, Wang, and Zhang}{Zhang
  et~al\mbox{.}}{2012}]%
        {zhang2012trust}
\bibfield{author}{\bibinfo{person}{Haibin Zhang}, \bibinfo{person}{Yan Wang},
  {and} \bibinfo{person}{Xiuzhen Zhang}.} \bibinfo{year}{2012}\natexlab{}.
\newblock \showarticletitle{A trust vector approach to transaction
  context-aware trust evaluation in e-commerce and e-service environments}. In
  \bibinfo{booktitle}{\emph{ICSOc}}. \bibinfo{pages}{1--8}.
\newblock


\bibitem[\protect\citeauthoryear{Zhang, Wang, Zhang, and et~al.}{Zhang
  et~al\mbox{.}}{2015}]%
        {zhang2015reputationpro}
\bibfield{author}{\bibinfo{person}{Haibin Zhang}, \bibinfo{person}{Yan Wang},
  \bibinfo{person}{Xiuzhen Zhang}, {and} \bibinfo{person}{et al.}}
  \bibinfo{year}{2015}\natexlab{}.
\newblock \showarticletitle{Reputationpro: the efficient approaches to
  contextual transaction trust computation in e-commerce environments}.
\newblock \bibinfo{journal}{\emph{ACM Transactions on the Web}}
  \bibinfo{volume}{9}, \bibinfo{number}{1} (\bibinfo{year}{2015}),
  \bibinfo{pages}{1--49}.
\newblock


\bibitem[\protect\citeauthoryear{Zhang, Wu, Yuan, Pan, and et~al.}{Zhang
  et~al\mbox{.}}{2022}]%
        {zhang2022trustworthy}
\bibfield{author}{\bibinfo{person}{He Zhang}, \bibinfo{person}{Bang Wu},
  \bibinfo{person}{Xingliang Yuan}, \bibinfo{person}{Shirui Pan}, {and}
  \bibinfo{person}{et al.}} \bibinfo{year}{2022}\natexlab{}.
\newblock \showarticletitle{Trustworthy graph neural networks: aspects, methods
  and trends}.
\newblock \bibinfo{journal}{\emph{arXiv preprint arXiv:2205.07424}}
  (\bibinfo{year}{2022}).
\newblock


\bibitem[\protect\citeauthoryear{Zhang, Yao, Sun, and Tay}{Zhang
  et~al\mbox{.}}{2019}]%
        {zhang2019deep}
\bibfield{author}{\bibinfo{person}{Shuai Zhang}, \bibinfo{person}{Lina Yao},
  \bibinfo{person}{Aixin Sun}, {and} \bibinfo{person}{Yi Tay}.}
  \bibinfo{year}{2019}\natexlab{}.
\newblock \showarticletitle{Deep learning based recommender system: a survey
  and new perspectives}.
\newblock \bibinfo{journal}{\emph{Comput. Surveys}} \bibinfo{volume}{52},
  \bibinfo{number}{1} (\bibinfo{year}{2019}), \bibinfo{pages}{1--38}.
\newblock


\bibitem[\protect\citeauthoryear{Zhang, Cui, and Wang}{Zhang
  et~al\mbox{.}}{2013}]%
        {zhang2013commtrust}
\bibfield{author}{\bibinfo{person}{Xiuzhen Zhang}, \bibinfo{person}{Lishan
  Cui}, {and} \bibinfo{person}{Yan Wang}.} \bibinfo{year}{2013}\natexlab{}.
\newblock \showarticletitle{Commtrust: computing multi-dimensional trust by
  mining e-commerce feedback comments}.
\newblock \bibinfo{journal}{\emph{IEEE Trans Knowl Data Eng}}
  \bibinfo{volume}{26}, \bibinfo{number}{7} (\bibinfo{year}{2013}),
  \bibinfo{pages}{1631--1643}.
\newblock


\bibitem[\protect\citeauthoryear{Zhang, Chen, et~al\mbox{.}}{Zhang
  et~al\mbox{.}}{2020}]%
        {zhang2020explainable}
\bibfield{author}{\bibinfo{person}{Yongfeng Zhang}, \bibinfo{person}{Xu Chen},
  {et~al\mbox{.}}} \bibinfo{year}{2020}\natexlab{}.
\newblock \showarticletitle{Explainable recommendation: a survey and new
  perspectives}.
\newblock \bibinfo{journal}{\emph{Foundations and Trends{\textregistered} in
  Information Retrieval}} \bibinfo{volume}{14}, \bibinfo{number}{1}
  (\bibinfo{year}{2020}), \bibinfo{pages}{1--101}.
\newblock


\bibitem[\protect\citeauthoryear{Zheng, Wang, Orgun, and et~al.}{Zheng
  et~al\mbox{.}}{2014}]%
        {zheng2014trust}
\bibfield{author}{\bibinfo{person}{Xiaoming Zheng}, \bibinfo{person}{Yan Wang},
  \bibinfo{person}{Mehmet Orgun}, {and} \bibinfo{person}{et al.}}
  \bibinfo{year}{2014}\natexlab{}.
\newblock \showarticletitle{Trust prediction with propagation and similarity
  regularization}. In \bibinfo{booktitle}{\emph{AAAI}}.
  \bibinfo{pages}{237--243}.
\newblock


\end{thebibliography}
